\documentclass{article}
\usepackage{arxiv}
\usepackage[utf8]{inputenc}
\usepackage{upgreek}
\usepackage{amsmath}
    \makeatletter
    \def\maketag@@@#1{\hbox{\m@th\normalfont\normalsize#1}}
    \makeatother
\usepackage{graphicx}
\usepackage{booktabs}
\usepackage{tabu}
\usepackage[T1]{fontenc}
\usepackage{subfigure}
\usepackage{siunitx} 
\usepackage{miller} 
\usepackage{soul}
\usepackage{fancyhdr}
\usepackage{tikz}
    \usetikzlibrary{calc,patterns,angles,quotes}
\usepackage{pgfplots}
    \pgfplotsset{compat=1.16}
\usetikzlibrary{decorations.pathreplacing}
\usepackage{float}
\usepackage{multirow}
\usepackage{fdsymbol}
\usepackage{mathtools}
\usepackage{leftindex}
\usepackage{amsfonts}
\usepackage{xfrac}
\usepackage{algorithm2e}
    \RestyleAlgo{ruled} 
    \SetKwInput{KwInput}{Input}
\usepackage[hidelinks]{hyperref}
    \hypersetup{
        colorlinks=false,
        pdftitle={A generalized calculation of the rate independent single crystal yield surface},
        pdfauthor={Matthew Kasemer},
    }
\usepackage{appendix}

\graphicspath{{figures/}}

\title{A generalized calculation of the rate independent single crystal yield surface}

\author{Matthew Kasemer \\
	Department of Mechanical Engineering\\
	The University of Alabama\\
	Tuscaloosa, AL 35487 \\
	\texttt{mkasemer@eng.ua.edu} \\
	\And
	Paul R. Dawson \\
	Sibley School of Mechanical and Aerospace Engineering\\
	Cornell University\\
	Ithaca, NY 14850 \\
	\texttt{prd5@cornell.edu} \\
}

\renewcommand{\shorttitle}{Kasemer et. al. 2025}

\hypersetup{
pdftitle={A generalized calculation of the rate independent single crystal yield surface},
pdfsubject={q-bio.NC, q-bio.QM},
pdfauthor={Matthew Kasemer, Paul Dawson},
pdfkeywords={Crystal, Yield, Yield surface, Crystal plasticity, Schmid's law},
}

\begin{document}
\maketitle

\begin{abstract}
In this paper, we discuss a method to calculate the topology of the rate independent single crystal yield surface for materials with arbitrary slip systems and arbitrary slip strengths. We describe the general problem, as motivated by Schmid's law, and detail the calculation of hyperplanes in deviatoric stress space, $\mathbb{D}^5$, which describe the criteria for slip on individual slip systems. We focus on finding the intersection of five linearly independent hyperplanes which represent stresses necessary to satisfy the criteria for general plastic deformation. Finally, we describe a method for calculating the inner convex hull of these intersection points, which describe the vertices of the five dimensional polytope that represents the single crystal yield surface. Our method applies to arbitrary crystal structure, allowing for an arbitrary number and type of slip systems and families, considers plastic anisotropy via inter- and intra-family strength anisotropy, and further considers strength anisotropy between slip in the positive and negative direction. We discuss the calculation and possible applications, and share a computational implementation of the calculation of the single crystal yield surface.
\end{abstract}

\keywords{Crystal \and
Yield \and
Yield surface \and
Crystal plasticity \and
Schmid's law}

\section{Introduction}
\label{sec:introduction}

Macroscopic stiffness and strength are perhaps the two most important material properties with regard to the design of metallic engineering components intended to both be subjected to mechanical loading and to operate within the elastic loading regime. In particular, the macroscopic strength, commonly referred to as the yield stress or simply yield, describes the stress limit of the elastic domain, and must be accurately quantified to assure that an engineering component functions within its expected operational window. Historically, such quantification has typically been achieved via experimental mechanical testing, though the establishment of mathematical models to describe the macroscopic yield behavior of materials has been the subject of significant interest since at least the work of Tresca in the mid 1800s~\cite{Tresca}, with notable development upon the work of von Mises in the 1910s~\cite{Mises1913}. Such mathematical frameworks offer the promise to predict the elastic limit of materials, offering engineers tools to employ in the design of engineering components.

The techniques of Tresca and von Mises, while powerful, inherently assume isotropic material behavior, and thus are representative of only a relatively small subset of materials that do not exhibit anisotropic yield behavior, thus necessitating further sophistication. The role of microstructure and the response of individual grains (or single crystals) on the macroscopic behavior of materials became evident by the early 1900s. For a variety of crystalline materials (including geological materials), it was established that a number of inelastic deformation mechanisms exist, such as slip, twinning/transformation induced plasticity, and diffusion. Overwhelmingly, however, we recognize that slip is the primary mechanism contributing to plastic deformation, and it has been observed in a wide range of loading environments. Following the work of von Mises~\cite{Mises1928} and Taylor~\cite{Taylor1934a,Taylor1934b}, the work of Schmid and Boas~\cite{Schmid1935} established the first rigorous mathematical description of the mechanics governing single crystal plasticity, and its role in the yield behavior of metals and metallic alloys. This behavior at the crystal scale is inherently anisotropic, and is dependent ultimately on crystal structure, alloying content, environment, and other considerations. Collectively---considering additionally the microstructural state (primarily crystallographic texture)---the plastic response of single crystals leads directly to anisotropic yield behavior exhibited at the macro scale, and development of understanding of the evolution of plasticity at the crystal scale is the cornerstone to understanding macroscopic deformation.

Bishop and Hill first formalized the concept of the single crystal yield surface, focusing on crystals exhibiting face centered cubic symmetry~\cite{Bishop1951}. This work established the single crystal yield surface as a surface in five dimensional deviatoric stress space, where points on the surface are stress states adequate to cause slip on at least one slip system. In other words, the surface itself represents the demarcation of the elastic regime of the single crystal, where the space enveloped by the single crystal yield surface represents all stress states that cause only elastic deformation. This work was further extended to crystals exhibiting body centered cubic symmetry deforming by pencil glide by Piehler and Backofen~\cite{Piehler1971}, and later to crystals exhibiting hexagonal close packed symmetry by Tom{\'e} and Kocks~\cite{Tome1985}. Collectively, these studies describe of the calculation of the single crystal yield surface for each of their respective problems (i.e., FCC and HCP crystals), though both studies employ a set of common assumptions and a strategy to define an irreducible volume in stress space. Chiefly, they rely on the persistence of crystal symmetry in the topology of the single crystal yield surface in deviatoric stress space. This assumption, in particular, eases the calculation of the single crystal yield surface, but implies that the single crystal is perfect (i.e., free of defect), as there is not any form of asymmetry in strength from slip system to slip system.

From the late 1980s onwards, development in deformation models at the polycrystalline scale (i.e., at a scale adequate to inform the macroscopic behavior of materials) have focused primarily on the formulation and employ of both viscoplastic self-consistent or full-field crystal plasticity finite element frameworks. The latter in particular---in both rate independent and rate dependent formulations---allows for high-fidelity predictions of the intragrain deformation response of polycrystalline aggregates and the establishment of microstructure-property relationships, though at a relatively high computational cost. Studies on the single crystal yield surface itself largely ebbed as focus turned toward evolving these frameworks to approach increasingly complex morphologies and phenomena. However, reevaluation of various assumptions made in the archival studies of the single crystal yield surface reveal that these prior results may be inadequate to address more complex behavior observed in the deformation response of materials. In particular, the heavy use of symmetry in these prior studies limits the results to only a subset of crystals: those which are perfect.

Consequently, we present in this study a more complete formalization of the general calculation of the rate-independent single crystal yield surface. We extend the formulations of the above studies to be entirely general, considering arbitrary crystal symmetry and allowing for asymmetry in the yield surface via inter- or intra-family strength anisotropy as well as strength anisotropy between the positive and negative slip direction. We contend that this tool helps to address complex behavior seen in materials which may no longer exhibit symmetry in slip behavior (e.g., irradiated materials, or latent hardening). Additionally, while this formulation has use even in modern crystal plasticity finite element frameworks (indeed, some full-field formulations continue to utilize the rate independent single crystal yield surface to address various needs), we contend that it has further use as an analytical tool to interpret the deformation response of materials. Finally, we attempt to simplify the description of the formulation and provide demonstrative figures to make the work of the archival publications more accessible, and additionally provide a software implementation disseminated free and open-source for public use and further development.

We begin this study by summarizing the theory of single crystal plasticity. Following, we describe our general methodology for determining the single crystal yield surface. We then present two demonstrations: the first considering planar two dimensional crystals (for ease of visualization), and the latter verification against archival publications (namely considering the results of both Bishop and Hill~\cite{Bishop1951} and Tom{\'e} and Kocks~\cite{Tome1985}). Finally, we present various extensions to more complex cubic and hexagonal systems not explored in previous literature before concluding.

\section{Single crystal plasticity}
\label{sec:schmid}

Crystallographic slip (or simply slip) is a phenomenological description of the complete traversal of dislocations through a material point by glide. If both the size of the material point and number of gliding dislocation are sufficiently large, then the discrete lattice displacements produced by dislocation traversal can be treated as continuous deformation. Typically, slip is characterized via the direction of dislocation motion, $\hat{d}_i$, (a normalized vector in the direction of the Burgers vector) on lattice planes with normals $\hat{n}_i$. This combination of slip direction and slip plane normal forms a slip system. A slip system is capable of producing a simple shear deformation when observed in a coordinate system aligned with the slip system directions. A collection of symmetrically equivalent slip systems forms a slip family. As dislocations with varying Burgers vector can move along different planes simultaneously, deformation can be accommodated by different slip systems simultaneously.

\begin{figure}[H]
    \centering
    \subfigure[]{%
        \label{subfig:sx_bcc}
        \resizebox{0.3\textwidth}{!}{
            \begin{tikzpicture}[scale=2] 
                \draw[thick,black,->,>=stealth](0,0,0)--(0,0,1.6)node[anchor=north east,inner sep=2pt]{$x$};
                \draw[thick,black,->,>=stealth](0,0,0)--(0,1.6,0)node[anchor=south,inner sep=2pt]{$z$};
                \draw[thick,black,->,>=stealth](0,0,0)--(1.6,0,0)node[anchor=west,inner sep=2pt]{$y$};
                \draw[thick,black](0,0,0)--(1,0,0)--(1,1,0)--(0,1,0)--cycle;
                \draw[thick,gray](0,0,0)--(1,0,0);
                
                \draw[thick,gray](0,0,0)--(0,.72,0);
                \draw[thick,black](0,0,1)--(1,0,1)--(1,1,1)--(0,1,1)--cycle;
                \draw[thick,black](0,0,0)--(0,0,1);
                \draw[thick,black](1,0,0)--(1,0,1);
                \draw[thick,black](1,1,0)--(1,1,1);
                \draw[thick,black](0,1,0)--(0,1,1);
                \draw[thick,gray](0,0,0)--(0,0,1);
                \filldraw[fill=gray,gray](0,0,0)circle(0.05);
                \filldraw[fill=black,black](1,0,0)circle(0.05);
                \filldraw[fill=black,black](0,1,0)circle(0.05);
                \filldraw[fill=black,black](0,0,1)circle(0.05);
                \filldraw[fill=black,black](1,1,0)circle(0.05);
                \filldraw[fill=black,black](1,0,1)circle(0.05);
                \filldraw[fill=black,black](0,1,1)circle(0.05);
                \filldraw[fill=black,black](1,1,1)circle(0.05);
                \filldraw[fill=black,black](0.5,0.5,0.5)circle(0.05);
                \draw [fill=gray,gray,opacity=0.2](1,0,0)--(1,1,0)--(0,1,1)--(0,0,1)--cycle;
                \draw[ultra thick,black,dashed](1,0,0)--(0,1,1);
                \draw[ultra thick,black,dashed](1,1,0)--(0,0,1);
            \end{tikzpicture}
        }
    }
    \subfigure[]{%
        \label{subfig:sx_fcc}
        \resizebox{0.3\textwidth}{!}{
            \begin{tikzpicture}[scale=2]
                \draw[thick,black,->,>=stealth](0,0,0)--(0,0,1.6)node[anchor=north east,inner sep=2pt]{$x$};
                \draw[thick,black,->,>=stealth](0,0,0)--(0,1.6,0)node[anchor=south,inner sep=2pt]{$z$};
                \draw[thick,black,->,>=stealth](0,0,0)--(1.6,0,0)node[anchor=west,inner sep=2pt]{$y$};
                \draw[thick,black](0,0,0)--(1,0,0)--(1,1,0)--(0,1,0)--cycle;
                \draw[thick,gray](.21,0,0)--(.72,0,0);
                \draw[thick,gray](0,.21,0)--(0,.72,0);
                \draw[thick,black](0,0,1)--(1,0,1)--(1,1,1)--(0,1,1)--cycle;
                \draw[thick,black](0,0,0)--(0,0,1);
                \draw[thick,black](1,0,0)--(1,0,1);
                \draw[thick,black](1,1,0)--(1,1,1);
                \draw[thick,black](0,1,0)--(0,1,1);
                \filldraw[fill=black,black](0,0,0)circle(0.05);
                \filldraw[fill=black,black](1,0,0)circle(0.05);
                \filldraw[fill=black,black](0,1,0)circle(0.05);
                \filldraw[fill=black,black](0,0,1)circle(0.05);
                \filldraw[fill=black,black](1,1,0)circle(0.05);
                \filldraw[fill=black,black](1,0,1)circle(0.05);
                \filldraw[fill=black,black](0,1,1)circle(0.05);
                \filldraw[fill=black,black](1,1,1)circle(0.05);
                \filldraw[fill=black,black](0.5,0.5,1)circle(0.05);
                \filldraw[fill=gray,gray](0.5,0.5,0)circle(0.05);
                \filldraw[fill=black,black](0.5,1,0.5)circle(0.05);
                \filldraw[fill=black,black](0.5,0,0.5)circle(0.05);
                \filldraw[fill=black,black](1,0.5,0.5)circle(0.05);
                \filldraw[fill=black,black](0,0.5,0.5)circle(0.05);
                \draw [fill=gray,gray,opacity=0.2](1,1,0)--(0,1,1)--(1,0,1)--cycle;
                \draw[ultra thick,black,dashed](1,0,1)--(0,1,1);
                \draw[ultra thick,black,dashed](1,0,1)--(1,1,0);
                \draw[ultra thick,black,dashed](0,1,1)--(1,1,0);
            \end{tikzpicture}
        }
    }
    \subfigure[]{%
        \label{subfig:sx_hcp}
        \resizebox{0.3\textwidth}{!}{
            \begin{tikzpicture}[scale=2] 
                \def\sfh{0.8}
                \draw[thick,black,->,>=stealth](.5*\sfh,0,.866*\sfh)--(.5*\sfh,0,.866*\sfh+1.6)node[anchor=north east,inner sep=2pt]{$x$};
                \draw[thick,black,->,>=stealth](.5*\sfh,0,.866*\sfh)--(.5*\sfh,0+1.6,.866*\sfh)node[anchor=south,inner sep=2pt]{$z$};
                \draw[thick,black,->,>=stealth](.5*\sfh,0,.866*\sfh)--(.5*\sfh+1.6,0,.866*\sfh)node[anchor=west,inner sep=2pt]{$y$};
                \draw[thick,gray](.5*\sfh,0,.866*\sfh)--(1.5*\sfh,0,.866*\sfh);
                \draw[thick,gray](.5*\sfh,0,.866*\sfh)--(.5*\sfh,1.385*\sfh,.866*\sfh);
                \draw[thick,gray](.5*\sfh,0,.866*\sfh)--(.5*\sfh,0,.866*\sfh+.09);
                \draw[thick,gray] (1*\sfh, 0, 0*\sfh) -- (1*\sfh, 1.387*\sfh, 0*\sfh);
                \draw[thick,black] 
                    (0*\sfh, 0, 0*\sfh) -- 
                    (1*\sfh, 0, 0*\sfh) -- 
                    (1.5*\sfh, 0, 0.866*\sfh) -- 
                    (1*\sfh, 0, 1.732*\sfh) -- 
                    (0*\sfh, 0, 1.732*\sfh) -- 
                    (-0.5*\sfh, 0, 0.866*\sfh) -- 
                    cycle;
                \draw[thick,black] 
                    (0*\sfh, 1.387*\sfh, 0*\sfh) -- 
                    (1*\sfh, 1.387*\sfh, 0*\sfh) -- 
                    (1.5*\sfh, 1.387*\sfh, 0.866*\sfh) -- 
                    (1*\sfh, 1.387*\sfh, 1.732*\sfh) -- 
                    (0*\sfh, 1.387*\sfh, 1.732*\sfh) -- 
                    (-0.5*\sfh, 1.387*\sfh, 0.866*\sfh) -- 
                    cycle;
                \foreach \x/\z in {0/0, 1.5/0.866, 1/1.732, 0/1.732, -0.5/0.866} {
                    \draw[thick,black] (\x*\sfh, 0, \z*\sfh) -- (\x*\sfh, 1.387*\sfh, \z*\sfh);
                }
                \draw[thick,gray] 
                    (0*\sfh, 0, 0*\sfh) -- 
                    (1*\sfh, 0, 0*\sfh);
                \draw[thick,gray] 
                    (1*\sfh, 0, 0*\sfh)--(1.5*\sfh, 0, 0.866*\sfh);
                \draw[thick,gray] 
                    (0*\sfh, 0, 0*\sfh) -- 
                    (-0.5*\sfh*.14, 0, 0.866*\sfh*.14);
                \draw[thick,gray] 
                    (0*\sfh, 0, 0*\sfh) -- 
                    (0*\sfh, 0.985*\sfh, 0*\sfh);
                \draw[thick,black] (1*\sfh, 0, 1.732*\sfh) -- (1*\sfh, 1.387*\sfh, 1.732*\sfh);
                \draw[thick,black] (1*\sfh, 1.387*\sfh, 1.732*\sfh) -- (0*\sfh, 1.387*\sfh, 1.732*\sfh);
                \foreach \x/\y/\z in {
                    0/0/0, 1/0/0, 1.5/0/0.866, 1/0/1.732, 0/0/1.732, -0.5/0/0.866, 
                    0/1.387/0, 1/1.387/0, 1.5/1.387/0.866, 1/1.387/1.732, 0/1.387/1.732, -0.5/1.387/0.866  
                } {
                    \filldraw[fill=black,black] (\x*\sfh, \y*\sfh, \z*\sfh) circle (0.05);
                }
                \foreach \x/\z in {0.5/0.866, 1/1.299, 0/1.299} {
                    \filldraw[fill=black,black] (\x*\sfh, 1.387/2, \z*\sfh) circle (0.05);
                }
                \filldraw[fill=gray,gray] (0.5*\sfh, 1.387/2, 0.866*\sfh) circle (0.05);
                \filldraw[fill=gray,gray] (0,0,0) circle (0.05);
                \filldraw[fill=gray,gray] (1*\sfh,0,0) circle (0.05);
                \draw [fill=gray,gray,opacity=0.2]
                    (0*\sfh, 0, 0*\sfh) -- 
                    (1*\sfh, 0, 0*\sfh) -- 
                    (1.5*\sfh, 0, 0.866*\sfh) -- 
                    (1*\sfh, 0, 1.732*\sfh) -- 
                    (0*\sfh, 0, 1.732*\sfh) -- 
                    (-0.5*\sfh, 0, 0.866*\sfh) -- 
                    cycle;
                \draw [fill=red,red,opacity=0.2]
                    (1*\sfh,0,1.732*\sfh) -- 
                    (1.5*\sfh,0,0.866*\sfh) -- 
                    (1.5*\sfh,1.386*\sfh,0.866*\sfh) -- 
                    (1*\sfh,1.386*\sfh,1.732*\sfh) -- 
                    cycle;
                \draw [fill=blue,blue,opacity=0.2]
                    (1*\sfh,0,1.732*\sfh) -- 
                    (1.5*\sfh,0,0.866*\sfh) -- 
                    (1*\sfh,1.386*\sfh,0) -- 
                    (0,1.386*\sfh,1.732*\sfh) -- 
                    cycle;
                \draw[ultra thick,black,dashed](1*\sfh,0,1.732*\sfh)--(1.5*\sfh,0,0.866*\sfh);
                \draw[ultra thick,black,loosely dashdotted](1*\sfh,0,1.732*\sfh)--(0,1.386*\sfh,1.732*\sfh);
                \draw[ultra thick,black,loosely dashdotted](1.5*\sfh,0,0.866*\sfh)--(1*\sfh,1.386*\sfh,0);
            \end{tikzpicture}
        }
    }
    \caption{Diagrams depicting slip planes and directions (dashed and dotted lines) for~\subref{subfig:sx_bcc} \hkl{110}\hkl<111> slip for body-centered cubic (BCC) symmetry,~\subref{subfig:sx_fcc} \hkl{111}\hkl<110> slip for face-centered cubic (FCC) symmetry, and~\subref{subfig:sx_hcp} \hkl{0001}\hkl<11-20> (gray plane, dashed line), \hkl{10-10}\hkl<11-20> (red plane, dashed line), and \hkl{10-11}\hkl<11-2-3> (blue plane, dash-dotted line) slip for hexagonal close-packed (HCP) symmetry.}
    \label{fig:sx}
\end{figure}
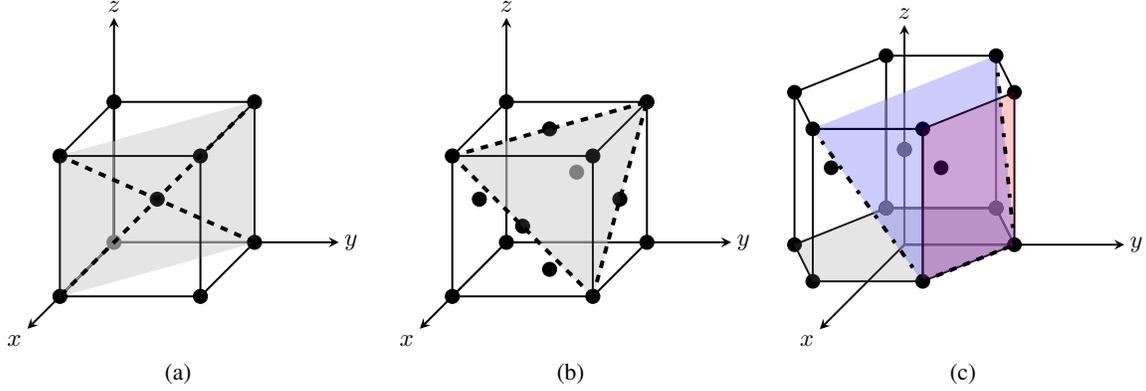

Slip in crystals is typically limited to a select number of slip systems, a phenomenon known as restricted slip. In crystals that exhibit cubic symmetry at room temperature, slip is commonly considered to be restricted to 12 slip systems from a single family---i.e., \hkl{110}\hkl<111> for body-centered cubic, or \hkl{111}\hkl<110> for face-centered cubic, which we illustrate in Figure~\ref{subfig:sx_bcc} and Figure~\ref{subfig:sx_fcc}, respectively. However, the slip systems that govern glide-based plasticity can become complex, especially with (for example, but not limited to) changes in temperature or lower crystal symmetry. In hexagonal close packed crystals, which exhibit lower symmetry than cubic crystals, slip is commonly considered to be restricted to 18 slip systems among three separate families (i.e., the basal \hkl{0001}\hkl<11-20>, prismatic \hkl{10-10}\hkl<11-20>, and pyramidal $c+a$ \hkl{10-11}\hkl<11-2-3> slip families), which we illustrate in Figure~\ref{subfig:sx_hcp}.

We now turn to the rigorous mathematical formalization of the mechanics governing slip in single crystals, as first formulated by~\cite{Schmid1935,Taylor1934a,Taylor1934b} and inspired heavily by~\cite{Bishop1951,Tome1985}. We note that all calculations herein are performed in a configuration of the crystal frame of reference---that is, a coordinate system aligned with the crystal coordinate system, as shown in Figure~\ref{fig:sx}. 

Schmid's law~\cite{Schmid1935} postulates that plastic deformation in crystals begins when the resolved shear stress, $\tau^s$, on a given slip system reaches a critical magnitude, $\tau$, or:
\begin{equation}
    \label{eq:rsseqcrss}
    \tau^s = \tau .
\end{equation}
To calculate the resolved shear stress on a slip system, we first turn to Cauchy's formula, which (generally) relates the traction on a surface, $t_i$, to the normal of that surface, $\hat{n}_j$:
\begin{equation}
    t_i = \sigma_{ji} \hat{n}_j .
\end{equation}
where the Cauchy stress, $\sigma_{ji}$, is a linear mapping between the vector space for the traction and the vector space for the plane normals. For the purposes of the plasticity of crystals (i.e., slip on a given slip system), we interpret $\hat{n}_j$ to be the plane normal for a slip system. To calculate the resolved shear stress on a given slip system, we find the magnitude of the component of the traction vector in the direction of slip on that plane, $\hat{d}_i$, via the vector inner product:
\begin{equation}
    \tau^s = t_i \hat{d}_i = \sigma_{ji} \hat{n}_j \hat{d}_i.
\end{equation}
The dyadic product between the slip plane normal and the slip direction is commonly referred to as the Schmid tensor, $T_{ji}$:
\begin{equation}
    \hat{n}_j \hat{d}_i = T_{ji}.
\end{equation}
Consequently, we write the calculation of the resolved shear stress as:
\begin{equation}
\label{eq:sx_yield_full}
    T_{ji} \sigma_{ji} = \tau^s,
\end{equation}
which we reduce further by recognizing that only the deviatoric and symmetric portions of both the Schmid tensor and Cauchy stress tensor contribute to the calculation of the resolved shear stress, since:
\begin{itemize}
    \item the Cauchy stress tensor is symmetric (i.e., $\sigma_{ji} = \sigma_{ij}$), owing to balance of angular momentum in the absence of body couples;
    \item the mean portion of the Cauchy stress produces no shear traction on any plane, meaning that only the deviatoric portion of the Cauchy stress (denoted $\sigma^\prime_{ij}$, itself symmetric) contributes to the calculation;
    \item the Schmid tensor is deviatoric, owing to the orthogonality of $\hat{n}$ and $\hat{d}$; and,
    \item the inner product of a skew tensor and a symmetric tensor is identically zero, meaning that only the symmetric portion of the Schmid tensor (denoted $P_{ij}$) contributes to the calculation, owing to the symmetry of the (deviatoric) Cauchy stress tensor.
\end{itemize}
We thus cast the calculation of the resolved shear stress (Equation~\ref{eq:sx_yield_full}) in a compact five-dimensional vector form, and further equate to the critical resolved shear stress (Equation~\ref{eq:rsseqcrss}), such that the criteria for slip is written as:
\begin{equation}
    \label{eq:schmidvec}
    p_i s_i = \tau,
\end{equation}
where $p_i$ and $s_i$ are deviatoric vectors constructed from their tensor components via:
\begin{equation}
    \label{eq:devschmidvec}
    p = \begin{Bmatrix}
            \sqrt{\frac{1}{2}} \left( P_{11} - P_{22} \right) \\
            \sqrt{\frac{3}{2}} P_{33} \\
            \sqrt{2} P_{23} \\
            \sqrt{2} P_{13} \\
            \sqrt{2} P_{12}
        \end{Bmatrix} ,
\end{equation}
and:
\begin{equation}
    \label{eq:svec}
    s = \begin{Bmatrix}
            \sqrt{\frac{1}{2}} \left( \sigma^\prime_{11} - \sigma^\prime_{22} \right) \\
            \sqrt{\frac{3}{2}} \sigma^\prime_{33} \\
            \sqrt{2} \sigma^\prime_{23} \\
            \sqrt{2} \sigma^\prime_{13} \\
            \sqrt{2} \sigma^\prime_{12}
        \end{Bmatrix} .
\end{equation}
We choose the scaling terms in these vectors such that vector inner product preserves the value obtained by the tensor inner product, or:
\begin{equation}
    P_{ij} \sigma_{ij}^\prime = p_i s_i .
\end{equation}

Ultimately, we interpret Equation~\ref{eq:schmidvec} as the equation for a hyperplane in $\mathbb{D}^5$ (or: five dimensional deviatoric stress space, where the basis is defined by unit tensorial components scaled via Equation~\ref{eq:svec}), where $p_i$ represents the normal of the hyperplane, and $\tau$ is the nearest distance of the hyperplane from the origin of deviatoric stress space. In other words, the hyperplane represents all of the deviatoric stress states, $s_i$, which will produce a resolved shear stress necessary to produce slip on that particular slip system. Points closer to the origin are deviatoric stress states that do not produce slip on that particular slip system, while points outside of the hyperplane are technically inadmissible and are only accessible via an increase to $\tau$, or hardening.

\section{Methodology for determining the single crystal yield surface for crystals with multiple slip system families and unequal slip system strengths}
\label{sec:methodology}

\subsection{Calculation of candidate stress vertices}
\label{subsec:cand_stress_verts}

We begin by recalling that the deviatoric component of stress is shown to contribute to the plastic deformation of metals (i.e., incompressible shape change), while the volumetric component contributes to the dilation or constriction of material (i.e., no shape change), and does not cause plastic deformation. Consequently, for general plastic deformation to occur---or to accommodate any arbitrary change in shape during plastic deformation (to satisfy an arbitrary deviatoric tensor)---five linearly independent slip systems must be active simultaneously~\cite{Taylor1938}. For this to occur, Equation~\ref{eq:schmidvec} must be satisfied for at least five different linearly independent slip systems simultaneously. Consequently, we seek to calculate the unique combinations of five linearly independent slip systems, from which we can ultimately identify the stress states which satisfy the conditions for generalized plastic deformation.

To begin, we find the total number of unique combinations of five slip systems, $M$, from a restricted set of $N$ slip systems to be $M=\leftindex[I]^N {C}_5$; this number includes all combinations of five slip systems, both linearly independent and linearly dependent combinations. Considering 12 available slip systems, for example, there exists a total of $M=\leftindex[I]^{12} {C}_5=792$ unique combinations of five slip systems, and for 18 available slip systems, there exists $M=\leftindex[I]^{18} {C}_5=8,568$ unique combinations of five slip systems. We test for necessary linear independence~\cite{Groves1963,Kocks1964} for any of the unique combinations of five slip systems (denoted here generically as slip systems $a$, $b$, $c$, $d$, and $e$), by constructing a matrix $\mathcal{P}_{ij}$~\cite{Tome1985}:
\begin{equation}
    \label{eq:schmidstens}
    \mathcal{P}_{ij} = \begin{bmatrix}
        p^a_j \\
        p^b_j \\
        p^c_j \\
        p^d_j \\
        p^e_j \\
    \end{bmatrix},
\end{equation}
where each row, $i$, is constructed of a deviatoric Schmid vector, $p_j$ (Equation~\ref{eq:devschmidvec}), related to one of the five slip systems in the set under consideration. Simply, for any unique set of five slip systems to consist of five linearly independent vectors, the determinant of $\mathcal{P}_{ij}$ must be non-zero. We construct $\mathcal{P}_{ij}$ and calculate its determinant for each of the $M$ combinations of five slip systems, and find the subset of these combinations which display linear independence (or non-zero determinants), with a total $M_L \leq M$.

Of these remaining combinations of linearly independent slip systems, for each set there must exist a point at which the five hyperplanes meet in $\mathbb{D}^5$ (a natural consequence of their linear independence). We interpret this point as a deviatoric stress which satisfies the condition set in Equation~\ref{eq:schmidvec} for all five slip systems simultaneously, or a stress state necessary to produce an arbitrary shape change. We find this point via the solution to the linear system:
\begin{equation}
    \label{eq:linearsystem}
    \mathcal{P}_{ij} s_j = \tau_i, 
\end{equation}
or, more completely:
\begin{equation}
    \label{eq:linearsystemfull}
    \begin{bmatrix}
        p^a_1 & p^a_2 & p^a_3 & p^a_4 & p^a_5 \\
        p^b_1 & p^b_2 & p^b_3 & p^b_4 & p^b_5 \\
        p^c_1 & p^c_2 & p^c_3 & p^c_4 & p^c_5 \\
        p^d_1 & p^d_2 & p^d_3 & p^d_4 & p^d_5 \\
        p^e_1 & p^e_2 & p^e_3 & p^e_4 & p^e_5 \\
    \end{bmatrix}
    \begin{Bmatrix}
        s_1 \\
        s_2 \\
        s_3 \\
        s_4 \\
        s_5
    \end{Bmatrix}
    = 
    \begin{Bmatrix}
        \tau^a \\
        \tau^b \\
        \tau^c \\
        \tau^d \\
        \tau^e
    \end{Bmatrix},
\end{equation}
where $s_j$ is the deviatoric stress state and $\tau_i$ is a vector of the critical resolved shear stresses for slip systems $a$, $b$, $c$, $d$, and $e$. The deviatoric stress state that solves this linear system of equation represents the vertex at the point where five hyperplanes meet, and will herein be referred to as a deviatoric stress vertex, or more simply a vertex.

In past studies by Bishop and Hill~\cite{Bishop1951} considering cubic crystal symmetry, and further by Tom{\'e} and Kocks~\cite{Tome1985} considering hexagonal crystal symmetry, various symmetry arguments were employed to efficiently find the set of vertices which comprise the inner convex hull. These symmetry arguments rely on the assumptions that:
\begin{enumerate}
    \item all slip systems in a given family share the same strength, and
    \item all slip systems have the same strength in both the positive and negative slip directions.
\end{enumerate}
In the case of Bishop and Hill's work, they considered slip on only one slip family with no presence of intra-family strength anisotropy or strength anisotropy between the positive and negative slip directions. The problem of inter-family strength anisotropy was confronted and addressed by Tom{\'e} and Kocks, who studied crystals exhibiting hexagonal symmetry, which display a propensity to slip on multiple slip families with unequal strengths. While Tom{\'e} and Kocks took great care to establish analytical formulations for the vertices of the single crystal yield surface as a function of the lattice ratio of hexagonal crystals, they too did not consider intra-family strength anisotropy or strength anisotropy between the positive and negative slip directions, and thus the above assumptions persist in facilitating advantageous yield surface symmetry in the case of hexagonal crystal symmetry, despite the presence of multiple slip families with inter-family strength anisotropy. 

Essentially, these previous studies assumed that the vector $\tau_i$ in Equation~\ref{eq:linearsystem} was limited to the case where all values from a single family had the same magnitude, as well as the same magnitude in the positive and negative slip direction, thus facilitating the employ of symmetry operators to infer the location of the full set of vertices. In this work, we consider the general case where the above assumptions may no longer be present. The consequence is that $\tau_i$ can potentially consider inter- and intra-family strength anisotropy, as well as the possibility of strength anisotropy between the positive and negative slip direction\footnote{Note that when considering polycrystalline materials, the orientation of the crystal is typically represented via any symmetrically equivalent orientation. In the case of different strengths in either positive or negative slip, the actual orientation must be used, as a symmetrically equivalent orientation may no longer maintain the proper strength magnitudes depending on the slip direction (i.e., some symmetrically equivalent orientations will flip these magnitudes).}. For this more general case, yield surface symmetry is potentially reduced, disallowing for the arguments of previous studies to be applied here to identify the vertices that comprise the inner convex hull. Operationally, owing to the consideration of intra-family strength anisotropy and/or the potential for strength anisotropy between the positive and negative slip directions, for a given combination of five linearly independent slip systems, Equation~\ref{eq:linearsystem} more broadly represents $2^5=32$ different linear systems of equations, considering the available permutations of $\pm \tau$ (an individually binary value) that can exist for the five individual components of the vector $\tau_i$. Solving all linear system of equations, we find a set of \emph{candidate} vertices, numbering $32M_L \geq M_C \geq M_L$.

\subsection{Identification of the single crystal yield surface}
\label{subsec:scys}

We seek to define a surface in $\mathbb{D}^5$ which describes the point at which a crystal will begin to slip based on an increase in deviatoric stress along any arbitrary direction in $\mathbb{D}^5$. This surface can be defined in multiple ways, though perhaps the most expedient and useful is via the set of vertices that define the intersection of the hyperplanes that define the surface. Since we have already calculated a set of candidate vertices, the challenge is finding the subset from this candidate set which comprise the {\emph{inner convex hull}} that ultimately represents the single crystal yield surface. We seek the inner convex hull for two reasons. First, we wish to find a {\emph{convex}} hull in an effort to satisfy Drucker's stability postulate, from which the consequence of yield surface convexity is established~\cite{Drucker1959}. Second, we wish to find the {\emph{inner}} convex hull as any convex hull which envelopes other candidate vertices would not accurately predict yield (i.e., it would envelope vertices/hyperplanes which themselves satisfy Schmid's law).

We opt for a brute-force method rather than reliance on the persistence of crystal symmetry in the topology of the single crystal yield surface to calculate the set of vertices which comprise the inner convex hull and thus represent the single crystal yield surface. We recognize that the points which belong to this set will lie either on or inside each of the hyperplanes found via Equation~\ref{eq:schmidvec}. Operationally, we can test this by comparing the location of each vertex relative to each of the hyperplanes, and make no assumptions regarding symmetry. Consequently, we define matrix, $h_{\alpha \beta}$, of size $\alpha = M_C$ (i.e., the number of candidate vertices) by $\beta = 2N$ (i.e., the total number of hyperplanes, considering both planes defined by the Schmid vector and its negative), which we note is (generally) not square. We calculate the components of the matrix via:
\begin{equation}
    \label{eq:hab}
    h_{\alpha \beta} = p^\beta_i\left(s_i^\alpha - \frac{\tau^\beta}{p_i^\beta}\right), 
\end{equation}
or the vector inner product between the normal of the hyperplane, $p_i^\beta$, and the difference vector between the vertex, $s_i^\alpha$, and a point on the hyperplane, $\sfrac{\tau^\beta}{p_i^\beta}$. In essence, this finds the component of the difference vector in the direction of the plane normal. If the dot product is positive, the stress vertex lies outside of the hyperplane (note: the angle associated with a positive dot product will be less than \SI{90}{\degree}). If the dot product is zero or negative, the stress vertex lies on or inside of the hyperplane, respectively (note: the angle associated with a negative or zero dot product will be equal to or greater than \SI{90}{\degree}, respectively).

If a row ($\alpha$) of $h_{\alpha \beta}$ is comprised of values that are all less than or equal to zero, then the candidate vertex in question lies on the inner convex hull, as it is either inside or on all possible hyperplanes. Conversely, if \emph{any} of the values on a row of $h_{\alpha \beta}$ are greater than zero, then the candidate vertex cannot lie on the inner convex hull, as it lies outside one (or more) of the hyperplanes. Thus, the rows with all values less than or equal to zero correspond to the candidate vertices that comprise the single crystal yield surface. This will yield a set of vertices which comprise the single crystal yield surface numbering $M_V \leq M_C$.

\subsection{Summarizing pseudo-algorithm}
\label{subsec:pseudo-algo}

Here, we present a pseudo-algorithm to describe the methodologies presented in Sections~\ref{subsec:cand_stress_verts} and~\ref{subsec:scys}. Additionally, a working computational implementation is shared with this document (see: Data Availability).

\begin{algorithm}[htbp!]
    \caption{Algorithm summarizing primary steps to determine the deviatoric stresses which describe the vertices of the single crystal yield surface in $\mathbb{D}^5$.}
    \label{alg:scys}
    \KwInput{$\hat{n}\left[3,N\right]$, $\hat{d}\left[3,N\right]$, $\tau\left[2,N\right]$}
    \KwResult{$s\left[5,M_V\right]$}
    Calculate $p = f\left(\hat{n},\hat{d}\right)$ for all $N$ slip systems \\
    Calculate all $M=\leftindex[I]^N {C}_5$ unique combinations of 5 slip systems \\
    \For{all $M$ combinations}{
        Construct $\mathcal{P} = f\left(p\right)$ \\
        {\If{$\det \mathcal{P} \neq 0$}{
            \For{all 32 permutations of $\{\pm\tau\}$}{
                Solve $\left[\mathcal{P}\right]\{s\}=\{\tau\}$ \\
                Increment $M_C$
                }
            }
        }
    }
    \For{$\alpha\gets1$ \KwTo $M_C$}{
        \For{$\beta\gets1$ \KwTo $2N$}{
            Calculate $h_{\alpha\beta}$
        }
        {\If{all($\beta\leq0$)}{
            Save $\{s\}^\alpha$ as a single crystal yield surface stress vertex \\
            Increment $M_V$
            }
        }
    }
\end{algorithm}

\section{Demonstration of the methodology for planar two dimensional crystals}
\label{sec:demonstration}

\subsection{Two slip systems}
\label{subsubsec:two_slip_systems}

To visualize the concepts presented in the previous sections is cumbersome owing to the high dimensionality and number of points under consideration. We instead here devote a brief section to a two dimensional case (i.e., $\mathbb{D}^2$), not to reflect any real problem, but simply to provide visualization of the concepts we employ in $\mathbb{D}^5$.

To begin, we initialize a problem at the point of Equation~\ref{eq:schmidvec}, i.e., where the equations for the hyperplanes are known. In this example, we assume that there are two slip systems (i.e., $N=2$), denoted $a$ and $b$, with deviatoric Schmid vectors defined as:
\begin{equation}
    \{p\}^{a} = \begin{Bmatrix}
        1 \\
        0
    \end{Bmatrix},
    \quad
    \{p\}^{b} = \begin{Bmatrix}
        0 \\
        1
    \end{Bmatrix},
\end{equation}
where $a$ and $b$ share the same slip system strength, which are further equal in magnitude in both the positive and negative slip directions, such that:
\begin{equation}
    \left|\tau^a\right|=\left|\tau^b\right|=1.
\end{equation}

In this case there exists only one possible combination of slip systems (i.e., $M=\leftindex[I]^2 {C}_2=1$): $a$ and $b$. Turning to Equation~\ref{eq:schmidstens}, we construct $\left[\mathcal{P}\right]$ via the individual deviatoric Schmid vectors:
\begin{equation}
    \begin{bmatrix}
        1 & 0 \\
        0 & 1
    \end{bmatrix},
\end{equation}
where $\left[\mathcal{P}\right]$ is defined only by two rows (Schmid vectors) owing to the reduced dimensionality of the problem (i.e., in $\mathbb{D}^2$, the problem becomes the intersection of two lines rather than 5 five dimensional hyperplanes as in $\mathbb{D}^5$). We note that the determinant of $\left[\mathcal{P}\right]$ is (by inspection) non-zero, thus demonstrating that slip systems $a$ and $b$ are linearly independent and their hyperplanes (or, in two dimensions: lines) are not parallel and must intersect (and thus, $M_L=M=1$).

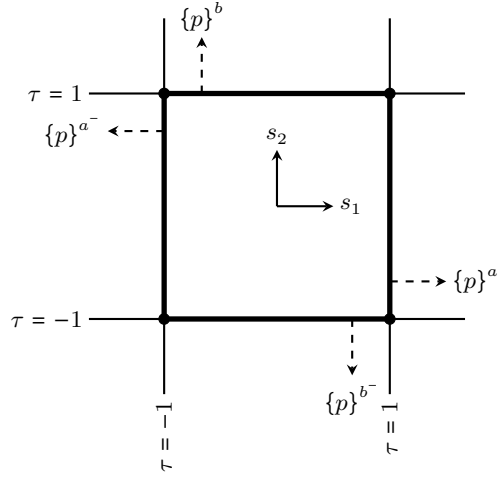
\begin{figure}[H]
    \centering
    \begin{tikzpicture}
        \draw[->,>=stealth,thick](0.0,0.0)--(0.75,0.0)node[anchor=west,inner sep=2pt]{\footnotesize$s_1$};
        \draw[->,>=stealth,thick](0.0,0.0)--(0.0,0.75)node[anchor=south,inner sep=2pt]{\footnotesize$s_2$};
        \draw[black,line width=2pt](-1.5,1.5)--(1.5,1.5);
        \draw[black,line width=2pt](1.5,1.5)--(1.5,-1.5);
        \draw[black,line width=2pt](1.5,-1.5)--(-1.5,-1.5);
        \draw[black,line width=2pt](-1.5,-1.5)--(-1.5,1.5);
        \draw[black,thick](-2.5,1.5)node[anchor=east,inner sep=2pt]{\footnotesize$\tau=1$}--(2.5,1.5);
        \draw[black,thick](-2.5,-1.5)node[anchor=east,inner sep=2pt]{\footnotesize$\tau=-1$}--(2.5,-1.5);
        \draw[black,thick](1.5,-2.5)node[anchor=north,inner sep=2pt]{\footnotesize\rotatebox{90}{$\tau=1$}}--(1.5,2.5);
        \draw[black,thick] (-1.5,-2.5)node[anchor=north,inner sep=2pt]{\footnotesize\rotatebox{90}{$\tau=-1$}}--(-1.5,2.5);
        \draw[->,>=stealth,dashed,thick](1.5,-1)--(2.25,-1)node[anchor=west,inner sep=2pt]{\footnotesize$\{p\}^a$};
        \draw[->,>=stealth,dashed,thick](-1.5,1)--(-2.25,1)node[anchor=east,inner sep=2pt]{\footnotesize$\{p\}^{a^-}$};
        \draw[->,>=stealth,dashed,thick](-1,1.5)--(-1,2.25)node[anchor=south,inner sep=2pt]{\footnotesize$\{p\}^b$};
        \draw[->,>=stealth,dashed,thick](1,-1.5)--(1,-2.25)node[anchor=north,inner sep=2pt]{\footnotesize$\{p\}^{b^-}$};
        \filldraw[black]( 1.5,1.5)circle(2pt)node[]{};
        \filldraw[black](-1.5,1.5)circle(2pt)node[]{};
        \filldraw[black]( 1.5,-1.5)circle(2pt)node[]{};
        \filldraw[black](-1.5,-1.5)circle(2pt)node[]{};
    \end{tikzpicture}
    \caption{Simplified two dimensional example of planes associated with slip conditions (solid lines) along with their normal vectors (dashed lines), and candidate vertices (dots) for the case of two possible slip systems---$a$ and $b$---plotted in deviatoric stress space. The single crystal yield surface is highlighted (thick solid lines).}
    \label{fig:2dys_1}
\end{figure}

We next construct the linear system of equations as defined by Equation~\ref{eq:linearsystem}:
\begin{equation}
    \begin{bmatrix}
        1 & 0 \\
        0 & 1
    \end{bmatrix}
    \{s\} =
    \{\tau\},
\end{equation}
recognizing that there exists $2^2=4$ positive/negative permutations of $\{\tau\}$ (and thus $M_C=4M_L=4$), or:
\begin{equation}
    \{\tau\} = 
    \begin{Bmatrix}
        \pm\tau^a\\
        \pm\tau^b
    \end{Bmatrix}
    =
    \begin{Bmatrix}
        1 \\
        1
    \end{Bmatrix},
    \quad
    \begin{Bmatrix}
        -1 \\
        1
    \end{Bmatrix},
    \quad
    \begin{Bmatrix}
        1 \\
        -1
    \end{Bmatrix},
    \quad
    \begin{Bmatrix}
        -1 \\
        -1
    \end{Bmatrix},
\end{equation}
for which the stress states, $\{s\}$, are trivially solved as $\left[\mathcal{P}\right]$ is identity. We interpret this to mean that due to slip on slip systems $a$ and $b$, there exists four possible deviatoric stress states which satisfy Schmid's law simultaneously for both slip systems (i.e., the von Mises criterion~\cite{Mises1928}) in either the positive or negative slip direction. We visualize this in Figure~\ref{fig:2dys_1}, and note that in this trivial example, the four candidate vertices are the same as the vertices that describe the inner convex hull (and thus $M_V = M_C = 4$).

\subsection{Three slip systems}
\label{subsubsec:three_slip_systems}

\begin{figure}[htbp!]
    \centering
    \begin{tikzpicture}
        \draw [->,>=stealth,thick](0.0,0.0)--(0.75,0.0)node[anchor=west,inner sep=2pt]{\footnotesize$s_1$};
        \draw [->,>=stealth,thick](0.0,0.0)--(0.0,0.75)node[anchor=south,inner sep=2pt]{\footnotesize$s_2$};
        \draw[black,line width=2pt](-1.5,1.5)--(1.1517,1.5);
        \draw[black,line width=2pt](1.1517,1.5)--(1.5,1.1517);
        \draw[black,line width=2pt](1.5,1.1517)--(1.5,-1.5);
        \draw[black,line width=2pt](1.5,-1.5)--(-1.1517,-1.5);
        \draw[black,line width=2pt](-1.1517,-1.5)--(-1.5,-1.1517);
        \draw[black,line width=2pt](-1.5,-1.1517)--(-1.5,1.5);
        \draw[black,thick](-4.5,1.5)node[anchor=east,inner sep=2pt]{\footnotesize$\tau=1$}--(4.5,1.5);
        \draw[black,thick](-4.5,-1.5)node[anchor=east,inner sep=2pt]{\footnotesize$\tau=-1$}--(4.5,-1.5);
        \draw[black,thick](1.5,-4.5)node[anchor=north,inner sep=2pt]{\rotatebox{90}{\footnotesize$\tau=1$}}--(1.5,4.5);
        \draw[black,thick](-1.5,-4.5)node[anchor=north,inner sep=2pt]{\rotatebox{90}{\footnotesize$\tau=-1$}}--(-1.5,4.5);
        \draw[black,thick](-1.8483,4.5)node[anchor=south east,inner sep=2pt]{\rotatebox{-45}{\footnotesize$\tau=1.25$}}--(4.5,-1.8483);
        \draw[black,thick](1.8483,-4.5)--(-4.5,1.8483)node[anchor=south east,inner sep=2pt]{\rotatebox{-45}{\footnotesize$\tau=-1.25$}};
        \draw [->,>=stealth,dashed,thick](1.5,-1)coordinate[](a1)--(2.25,-1)coordinate[](a2) node[anchor=west,inner sep=2pt]{\footnotesize$\{p\}^a$};
        \draw [->,>=stealth,dashed,thick](-1.5,1)coordinate[](a1m)--(-2.25,1)coordinate[](a2m) node[anchor=east,inner sep=2pt]{\footnotesize$\{p\}^{a^-}$};
        \draw [->,>=stealth,dashed,thick](-1,1.5)coordinate[](b1)--(-1,2.25)coordinate[](b2) node[anchor=south,inner sep=2pt]{\footnotesize$\{p\}^b$};
        \draw [->,>=stealth,dashed,thick](1,-1.5)coordinate[](b1m)--(1,-2.25)coordinate[](b2m) node[anchor=north,inner sep=2pt]{\footnotesize$\{p\}^{b^-}$};
        \draw [->,>=stealth,dashed,thick](2.6517,0)coordinate[](c1)--(3.182,0.53033)coordinate[](c2)node[anchor=south west,inner sep=2pt]{\footnotesize$\{p\}^c$};
        \draw [->,>=stealth,dashed,thick](-2.6517,0)coordinate[](c1m)--(-3.182,-0.53033)coordinate[](c2m)node[anchor=north east,inner sep=2pt]{\footnotesize$\{p\}^{c^-}$};
        \filldraw[black](1.5,1.5)circle(2pt)node[anchor=south west,inner sep=2pt]{\footnotesize$1$};
        \filldraw[black](-1.5,1.5)circle(2pt)node[anchor=south east,inner sep=2pt]{\footnotesize$2$};
        \filldraw[black](-1.5,-1.5)circle(2pt)node[anchor=north east,inner sep=2pt]{\footnotesize$3$};
        \filldraw[black](1.5,-1.5)circle(2pt)node[anchor=north west,inner sep=2pt]{\footnotesize$4$};
        \filldraw[black](1.5,1.1517)circle(2pt)node[anchor=north east,inner sep=2pt]{\footnotesize$5$};
        \filldraw[black](1.1517,1.5)circle(2pt)node[anchor=north east,inner sep=2pt]{\footnotesize$6$};
        \filldraw[black](-1.5,-1.1517)circle(2pt)node[anchor=south west,inner sep=2pt]{\footnotesize$7$};
        \filldraw[black](-1.1517,-1.5)circle(2pt)node[anchor=south west,inner sep=2pt]{\footnotesize$8$};
        \filldraw[black](4.1517,-1.5)circle(2pt)node[anchor=south west,inner sep=2pt]{\footnotesize$9$};
        \filldraw[black](-1.5,4.1517)circle(2pt)node[anchor=south west,inner sep=2pt]{\footnotesize$10$};
        \filldraw[black](-4.1517,1.5)circle(2pt)node[anchor=south west,inner sep=2pt]{\footnotesize$11$};
        \filldraw[black](1.5,-4.1517)circle(2pt)node[anchor=south west,inner sep=2pt]{\footnotesize$12$};
    \end{tikzpicture}
    \caption{Simplified two dimensional example of planes associated with slip conditions (solid lines) along with their normal vectors (dashed lines), and candidate vertices (dots, numbered) for the case of three possible slip systems---$a$, $b$, and $c$---plotted in deviatoric stress space. The single crystal yield surface is highlighted (thick solid lines).}
    \label{fig:2dys_2}
\end{figure}
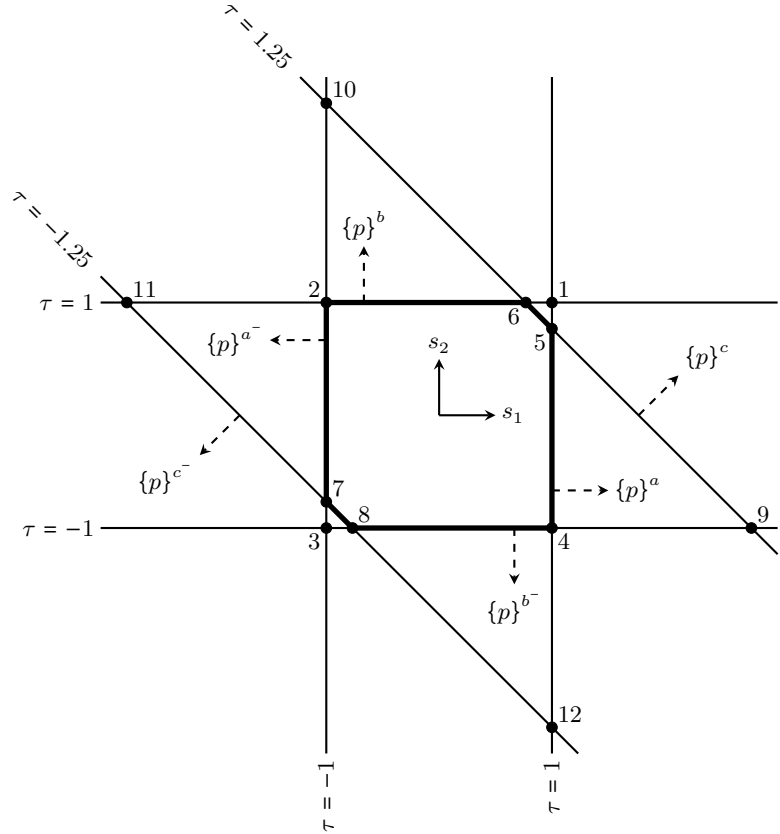

To provide a more interesting example for the case of calculating the single crystal yield surface, we present a modification and introduce an additional slip system, $c$, defined by the Schmid vector:
\begin{equation}
    \{p\}^{c} = \begin{Bmatrix}
        \frac{\sqrt{2}}{2} \\
        \frac{\sqrt{2}}{2}
    \end{Bmatrix}
\end{equation}
with a strength equal in both positive and negative strength of magnitude:
\begin{equation}
    \left|\tau^c\right| = 1.25.
\end{equation}
In this case there exists three possible combination of slip systems (i.e., $M=\leftindex[I]^3 {C}_2=3$, or the combinations of slip systems $a$ and $b$, $a$ and $c$, and $b$ and $c$) and of those combinations, all are linearly independent ($M_L=3$), as any two non-parallel lines in two dimensions are linearly independent (i.e., we do not need to belabor the point made by the determinant of the matrix in Equation~\ref{eq:schmidstens} for this simple example). We further recognize that there again exists $2^2=4$ permutations of $\{\pm\tau\}$ for each of the 3 combinations of linearly independent slip systems, and thus the number of candidate vertices numbers $M_C=4M_L=12$. We visualize the hyperplanes and candidate vertices in Figure~\ref{fig:2dys_2}.

The question now turns to which of these 12 vertices belong to the inner convex hull and thus (collectively) describe the single crystal yield surface. In this simplified example, the answer is again evident by inspection: vertices 2, 4, 5, 6, 7, and 8 comprise the inner convex hull (and thus $M_V = 6$). We further note that in this example the symmetry of the single crystal yield surface is reduced relative to that exhibited in Figure~\ref{fig:2dys_1}. Despite the relative ease of visually identifying the solution, this example provides a clear demonstration in which there exists a \emph{cloud} of candidate vertices from which we must determine which belong to the inner convex hull. 

\begin{figure}[H]
    \centering
    \subfigure[]{%
        \label{subfig:2dys_3_a}
        \resizebox{0.48\textwidth}{!}{
        \begin{tikzpicture}
            \draw[black,thick](-2.5,1.5)--(2.5,1.5);
            \draw[black,thick](-2.5,-1.5)--(2.5,-1.5);
            \draw[black,thick](1.5,-2.5)--(1.5,2.5);
            \draw[black,thick](-1.5,-2.5)--(-1.5,2.5);
            \draw[black,thick](-0.34835,3)--(3,-0.34835);
            \draw[black,thick](0.34835,-3)--(-3,0.34835);
            \filldraw[black](1.5,1.5)circle(2pt)node[anchor=south west,inner sep=2pt]{\footnotesize$1$};
            \filldraw[black](-1.5,1.5)circle(2pt)node[anchor=south east,inner sep=2pt]{\footnotesize$2$};
            \filldraw[black](-1.5,-1.5)circle(2pt)node[anchor=north east,inner sep=2pt]{\footnotesize$3$};
            \filldraw[black](1.5,-1.5)circle(2pt)node[anchor=north west,inner sep=2pt]{\footnotesize$4$};
            \filldraw[black](1.5,1.1517)circle(2pt)node[anchor=north east,inner sep=2pt]{\footnotesize$5$};
            \filldraw[black](1.1517,1.5)circle(2pt)node[anchor=north east,inner sep=2pt]{\footnotesize$6$};
            \filldraw[black](-1.5,-1.1517)circle(2pt)node[anchor=south west,inner sep=2pt]{\footnotesize$7$};
            \filldraw[black](-1.1517,-1.5)circle(2pt)node[anchor=south west,inner sep=2pt]{\footnotesize$8$};
            \draw [->,>=stealth,dashed,thick](1.5,-1)coordinate[](a1)--(2.25,-1)coordinate[](a2) node[anchor=west,inner sep=2pt]{\footnotesize$\{p\}^a$};
            \draw [->,>=stealth,dashed,thick](-1.5,1)coordinate[](a1m)--(-2.25,1)coordinate[](a2m) node[anchor=east,inner sep=2pt]{\footnotesize$\{p\}^{a^-}$};
            \draw [->,>=stealth,dashed,thick](-1,1.5)coordinate[](b1)--(-1,2.25)coordinate[](b2) node[anchor=south,inner sep=2pt]{\footnotesize$\{p\}^b$};
            \draw [->,>=stealth,dashed,thick](1,-1.5)coordinate[](b1m)--(1,-2.25)coordinate[](b2m) node[anchor=north,inner sep=2pt]{\footnotesize$\{p\}^{b^-}$};
            \draw [->,>=stealth,dashed,thick](2.6517,0)coordinate[](c1)--(3.182,0.53033)coordinate[](c2)node[anchor=south west,inner sep=2pt]{\footnotesize$\{p\}^c$};
            \draw [->,>=stealth,dashed,thick](-2.6517,0)coordinate[](c1m)--(-3.182,-0.53033)coordinate[](c2m)node[anchor=north east,inner sep=2pt]{\footnotesize$\{p\}^{c^-}$};
            \draw[black,dotted,thick](1.5,1.5)coordinate[](O)--(1.5,-0.5);
            \draw[black,dotted,thick](1.5,1.5)--(-1.5,1);
            \draw[black,dotted,thick](1.5,1.5)--(-1,1.5);
            \draw[black,dotted,thick](1.5,1.5)--(1,-1.5);
            \draw[black,dotted,thick](1.5,1.5)--(2.6517,0);
            \draw[black,dotted,thick](1.5,1.5)--(-2.6517,0);
            \pic[draw,<->,angle radius=3mm]{angle=a2--a1--O};
            \pic[draw,<->,angle radius=3mm]{angle=O--a1m--a2m};
            \pic[draw,<->,angle radius=3mm]{angle=O--b1--b2};
            \pic[draw,<->,angle radius=3mm]{angle=b2m--b1m--O};
            \pic[draw,<->,angle radius=3mm]{angle=c2--c1--O};
            \pic[draw,<->,angle radius=3mm]{angle=c2m--c1m--O};  
        \end{tikzpicture}
        }
    }
    \subfigure[]{%
        \label{subfig:2dys_3_b}
        \resizebox{0.48\textwidth}{!}{
        \begin{tikzpicture}
            \draw[black,thick](-2.5,1.5)--(2.5,1.5);
            \draw[black,thick](-2.5,-1.5)--(2.5,-1.5);
            \draw[black,thick](1.5,-2.5)--(1.5,2.5);
            \draw[black,thick](-1.5,-2.5)--(-1.5,2.5);
            \draw[black,thick](-0.34835,3)--(3,-0.34835);
            \draw[black,thick](0.34835,-3)--(-3,0.34835);
            \filldraw[black](1.5,1.5)circle(2pt)node[anchor=south west,inner sep=2pt]{\footnotesize$1$};
            \filldraw[black](-1.5,1.5)circle(2pt)node[anchor=south east,inner sep=2pt]{\footnotesize$2$};
            \filldraw[black](-1.5,-1.5)circle(2pt)node[anchor=north east,inner sep=2pt]{\footnotesize$3$};
            \filldraw[black](1.5,-1.5)circle(2pt)node[anchor=north west,inner sep=2pt]{\footnotesize$4$};
            \filldraw[black](1.5,1.1517)circle(2pt)node[anchor=north east,inner sep=2pt]{\footnotesize$5$};
            \filldraw[black](1.1517,1.5)circle(2pt)node[anchor=north east,inner sep=2pt]{\footnotesize$6$};
            \filldraw[black](-1.5,-1.1517)circle(2pt)node[anchor=south west,inner sep=2pt]{\footnotesize$7$};
            \filldraw[black](-1.1517,-1.5)circle(2pt)node[anchor=south west,inner sep=2pt]{\footnotesize$8$};
            \draw [->,>=stealth,dashed,thick](1.5,-1)coordinate[](a1)--(2.25,-1)coordinate[](a2) node[anchor=west,inner sep=2pt]{\footnotesize$\{p\}^a$};
            \draw [->,>=stealth,dashed,thick](-1.5,1)coordinate[](a1m)--(-2.25,1)coordinate[](a2m) node[anchor=east,inner sep=2pt]{\footnotesize$\{p\}^{a^-}$};
            \draw [->,>=stealth,dashed,thick](-1,1.5)coordinate[](b1)--(-1,2.25)coordinate[](b2) node[anchor=south,inner sep=2pt]{\footnotesize$\{p\}^b$};
            \draw [->,>=stealth,dashed,thick](1,-1.5)coordinate[](b1m)--(1,-2.25)coordinate[](b2m) node[anchor=north,inner sep=2pt]{\footnotesize$\{p\}^{b^-}$};
            \draw [->,>=stealth,dashed,thick](2.6517,0)coordinate[](c1)--(3.182,0.53033)coordinate[](c2)node[anchor=south west,inner sep=2pt]{\footnotesize$\{p\}^c$};
            \draw [->,>=stealth,dashed,thick](-2.6517,0)coordinate[](c1m)--(-3.182,-0.53033)coordinate[](c2m)node[anchor=north east,inner sep=2pt]{\footnotesize$\{p\}^{c^-}$};
            \draw[black,dotted,thick](1.5,1.1517)coordinate[](O)--(1.5,-0.5);
            \draw[black,dotted,thick](1.5,1.1517)--(-1.5,1);
            \draw[black,dotted,thick](1.5,1.1517)--(-1,1.5);
            \draw[black,dotted,thick](1.5,1.1517)--(1,-1.5);
            \draw[black,dotted,thick](1.5,1.1517)--(2.6517,0);
            \draw[black,dotted,thick](1.5,1.1517)--(-2.6517,0);
            \pic[draw,<->,angle radius=3mm]{angle=a2--a1--O};
            \pic[draw,<->,angle radius=3mm]{angle=O--a1m--a2m};
            \pic[draw,<->,angle radius=3mm]{angle=O--b1--b2};
            \pic[draw,<->,angle radius=3mm]{angle=b2m--b1m--O};
            \pic[draw,<->,angle radius=3mm]{angle=c2--c1--O};
            \pic[draw,<->,angle radius=3mm]{angle=c2m--c1m--O};  
        \end{tikzpicture}
        }
    }
    \caption{Example calculation of the values for a row of $h_{\alpha\beta}$ for the simplified two dimensional example presented in Figure~\ref{fig:2dys_2}, depicting the planes associated with slip conditions (solid lines) along with their normal vectors (dashed lines), and candidate vertices (dots, numbered) for the case of three possible slip systems---$a$, $b$, and $c$---plotted in deviatoric stress space. The difference vectors between points on each hyperplane and a single stress vertex are shown (dotted lines), along with the angle between the difference vectors and the plane normals. Examples shown for \subref{subfig:2dys_3_a} candidate vertex 1, a vertex that does not reside on the single crystal yield surface, and \subref{subfig:2dys_3_b} candidate vertex 5, a vertex which does reside on the single crystal yield surface. Candidate vertices 9, 10, 11, and 12, as well as the stress-space coordinate axes, are not shown for sake of simplicity.}
    \label{fig:2dys_3}
\end{figure}
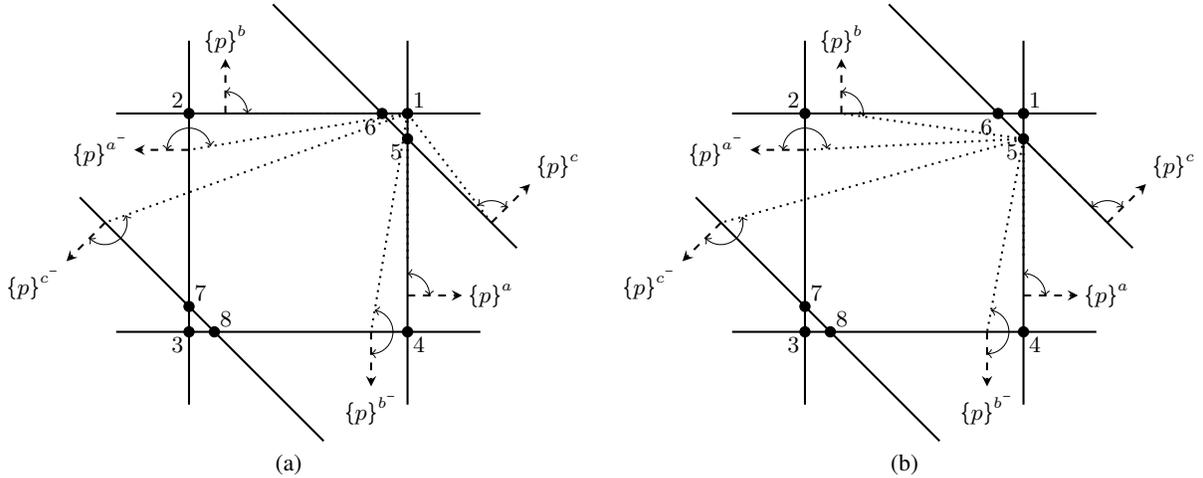

Consequently, we employ the method presented in Section~\ref{subsec:scys} here to calculate which vertices belong to the single crystal yield surface. To visualize, Figure~\ref{fig:2dys_3} shows the same example as shown in Figure~\ref{fig:2dys_2}, though with a demonstration of the calculation of the components of $h_{\alpha\beta}$ for two different candidate vertices: 1 and 5. Figure~\ref{fig:2dys_3} provides a depiction of the angle between a hyperplane normal (i.e., $p_i^\beta$ in Equation~\ref{eq:hab}) and the difference vector between an arbitrary point on the hyperplane and a candidate vertex (i.e., the term in the parentheses in Equation~\ref{eq:hab}), from which we can infer the value of $h_{\alpha\beta}$ (refer to the discussion in Section~\ref{subsec:scys}). In Figure~\ref{subfig:2dys_3_a}, we note that some angles are less than \SI{90}{\degree}, and thus vertex in question lies outside some hyperplanes, confirming that it does not belong to the single crystal yield surface. In Figure~\ref{subfig:2dys_3_b}, we note that all angles are equal to or greater than \SI{90}{\degree}, and thus vertex lies on or inside all hyperplanes, respectively, confirming that it does belong to the single crystal yield surface.

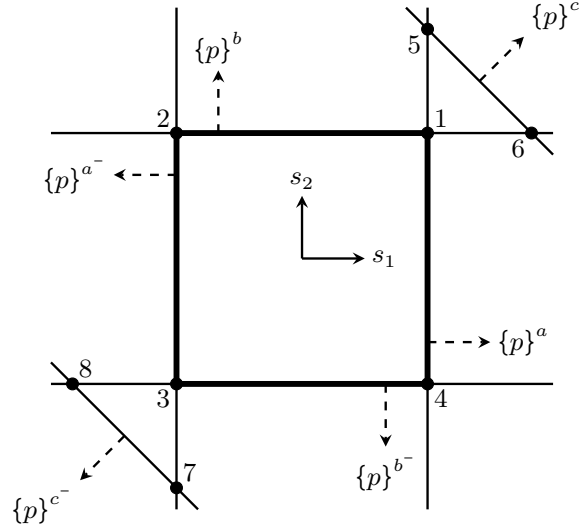
\begin{figure}[htbp!]
    \centering
    \resizebox{0.48\textwidth}{!}{
    \begin{tikzpicture}
        \draw[->,>=stealth,thick](0.0,0.0)--(0.75,0.0)node[anchor=west,inner sep=2pt]{\footnotesize$s_1$};
        \draw[->,>=stealth,thick](0.0,0.0)--(0.0,0.75)node[anchor=south,inner sep=2pt]{\footnotesize$s_2$};
        \draw[black,thick](-3,1.5)--(3,1.5);
        \draw[black,thick](-3,-1.5)--(3,-1.5);
        \draw[black,thick](1.5,-3)--(1.5,3);
        \draw[black,thick](-1.5,-3)--(-1.5,3);
        \draw[black,thick](1.2426,3)--(3,1.2426);
        \draw[black,thick](-1.2426,-3)--(-3,-1.2426);
        \draw[black,line width=2pt](-1.5,1.5)--(1.5,1.5);
        \draw[black,line width=2pt](1.5,1.5)--(1.5,-1.5);
        \draw[black,line width=2pt](1.5,-1.5)--(-1.5,-1.5);
        \draw[black,line width=2pt](-1.5,-1.5)--(-1.5,1.5);
        \filldraw[black](1.5,1.5)circle(2pt)node[anchor=south west,inner sep=2pt]{\footnotesize$1$};
        \filldraw[black](-1.5,1.5)circle(2pt)node[anchor=south east,inner sep=2pt]{\footnotesize$2$};
        \filldraw[black](-1.5,-1.5)circle(2pt)node[anchor=north east,inner sep=2pt]{\footnotesize$3$};
        \filldraw[black](1.5,-1.5)circle(2pt)node[anchor=north west,inner sep=2pt]{\footnotesize$4$};
        \filldraw[black](1.5,2.7426)circle(2pt)node[anchor=north east,inner sep=2pt]{\footnotesize$5$};
        \filldraw[black](2.7426,1.5)circle(2pt)node[anchor=north east,inner sep=2pt]{\footnotesize$6$};
        \filldraw[black](-1.5,-2.7426)circle(2pt)node[anchor=south west,inner sep=2pt]{\footnotesize$7$};
        \filldraw[black](-2.7426,-1.5)circle(2pt)node[anchor=south west,inner sep=2pt]{\footnotesize$8$};
        \draw [->,>=stealth,dashed,thick](1.5,-1)coordinate[](a1)--(2.25,-1)coordinate[](a2) node[anchor=west,inner sep=2pt]{\footnotesize$\{p\}^a$};
        \draw [->,>=stealth,dashed,thick](-1.5,1)coordinate[](a1m)--(-2.25,1)coordinate[](a2m) node[anchor=east,inner sep=2pt]{\footnotesize$\{p\}^{a^-}$};
        \draw [->,>=stealth,dashed,thick](-1,1.5)coordinate[](b1)--(-1,2.25)coordinate[](b2) node[anchor=south,inner sep=2pt]{\footnotesize$\{p\}^b$};
        \draw [->,>=stealth,dashed,thick](1,-1.5)coordinate[](b1m)--(1,-2.25)coordinate[](b2m) node[anchor=north,inner sep=2pt]{\footnotesize$\{p\}^{b^-}$};
        \draw [->,>=stealth,dashed,thick](2.1213,2.1213)coordinate[](c1)--(2.6517,2.6517)coordinate[](c2)node[anchor=south west,inner sep=2pt]{\footnotesize$\{p\}^c$};
        \draw [->,>=stealth,dashed,thick](-2.1213,-2.1213)coordinate[](c1m)--(-2.6517,-2.6517)coordinate[](c2m)node[anchor=north east,inner sep=2pt]{\footnotesize$\{p\}^{c^-}$}; 
    \end{tikzpicture}
    }
    \caption{Simplified two dimensional example of planes associated with slip conditions (solid lines) along with their normal vectors (dashed lines), and candidate vertices (dots, numbered) for the case of three possible slip systems---$a$, $b$, and $c$---plotted in deviatoric stress space. In this example, the critical resolved shear stress of slip system $c$ is to demonstrate the effects of relatively high slip system strength on the single crystal yield surface calculation. The single crystal yield surface is highlighted (thick solid lines). Candidate vertices 9, 10, 11, and 12 are not shown for sake of simplicity}
    \label{fig:2dys_4}
\end{figure}

When considering all three slip systems, if the critical resolved shear stress for slip system $c$ is increased beyond a critical point, vertices 5, 6, 7, and 8 will fall outside of the hyperplanes associated with slip on slip systems $a$ and $b$. Particularly, if the critical resolved shear stress, $\left|\tau^c\right|$, is greater than $\sqrt{2}$, then slip cannot occur on slip system $c$, and the single crystal yield surface reverts back to the four vertices shown in Figure~\ref{fig:2dys_1}. We present a depiction of this in Figure~\ref{fig:2dys_4}, for the case of $\left|\tau^c\right| = 2 \ge \sqrt{2}$, and note how the single crystal yield surface indeed reverts back to that comprised of only slip systems $a$ and $b$ (and $M_V=4$), demonstrating how the relatively high strength of some slip systems may render their contribution to the single crystal yield surface null. 

\begin{figure}[htbp!]
    \centering
    \includegraphics[width=3.5in]{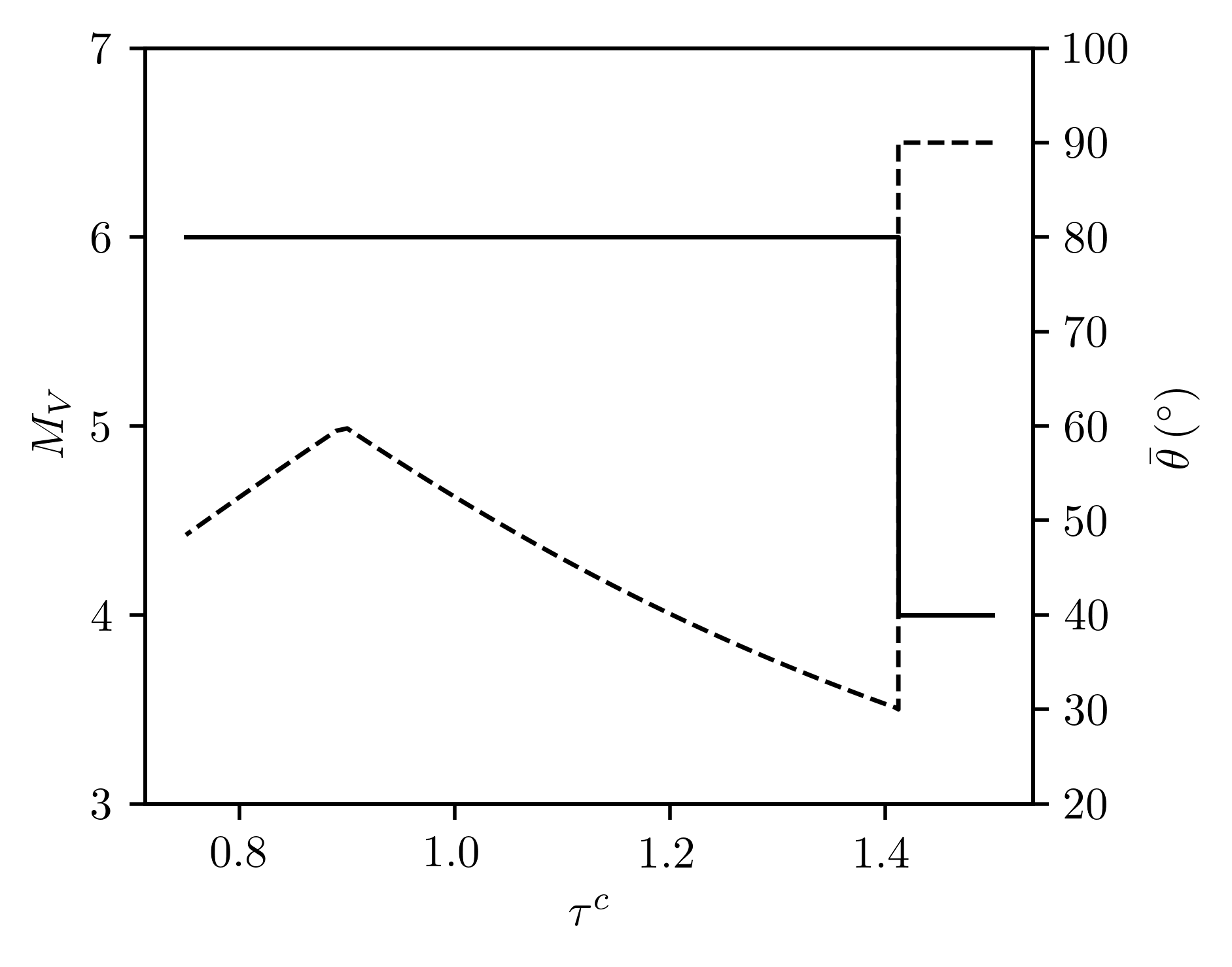}
    \caption{Evolution of the number of vertices, $M_V$, (solid line, left vertical axis) and the average angle to the nearest neighbor, $\bar{\theta}$, (dashed line, right vertical axis) as a function of $\tau^c$.}
    \label{fig:2d_trends}
\end{figure}

Per the examples shown in Figures~\ref{fig:2dys_2} and~\ref{fig:2dys_4}, we pause here to discuss the behavior regarding the spacing of the single crystal yield surface vertices as a function of the strength of slip system $c$ relative to the strength of slip systems $a$ and $b$. To begin, we define a metric, $\bar{\theta}$:
\begin{equation}
    \label{eq:bartheta}
    \bar{\theta} = \frac{1}{M_V}\sum_{a}^{M_V}{\arccos{\left(\frac{s_i^v s_i^w}{\left|s_i^v\right|\left|s_i^w\right|}\right)}},
\end{equation}
or the average over all single crystal yield surface vertices of the angle between a vertex, $s^v_i$, and its nearest neighbor, $s_i^w$. For the example considering all three slip systems, we plot the behavior of $\bar{\theta}$ along with the number of single crystal yield surface vertices, $M_V$, as a function of $\tau^c$ in Figure~\ref{fig:2d_trends}.

When $\left|\tau^c\right|>\sqrt{2}$ and the single crystal yield surface is comprised of vertices 1, 2, 3, and 4, the angle between each vertex and its nearest neighbor is \SI{90}{\degree} (and thus the average over all vertices is \SI{90}{\degree}). When $\left|\tau^c\right|$ is marginally below $\sqrt{2}$, vertices 5 and 6 converge to nearly the same point (similarly for vertices 7 and 8)---i.e., the angles between these vertices and their nearest neighbors are nearly zero while the angle between vertices 2 and 4 and their nearest neighbors is nearly \SI{90}{\degree}, and thus the average over all vertices is approximately \SI{30}{\degree}. As the strength of slip system $c$ is further decreased below $\sqrt{2}$, the angle to the nearest neighbor for vertices 5 and 6 (and 7 and 8) will increase while the angle to the nearest neighbor for vertices 2 and 4 will decrease, and overall the average increases. At approximately $\tau^c=0.9$, vertices 6 and 7 will be closer to vertex 2 than to vertices 5 and 8 respectively (similarly, vertices 5 and 8 will be closer to vertex 4 than to vertices 6 and 7), and the average angle to the nearest neighbor will begin to decrease as $\tau^c$ is decreased further. Overall, we observe that the inclusion of slip system $c$ and the value of its relative strength has an effect on both the number of vertices on the single crystal yield surface, $M_V$, and the average spacing of vertices and their nearest neighbor in deviatoric stress space, $\bar{\theta}$, both as a function of the relative strength between slip systems.

\begin{figure}[H]
    \centering
    \resizebox{0.48\textwidth}{!}{
    \begin{tikzpicture}
        \draw[->,>=stealth,thick](0.0,0.0)--(0.75,0.0)node[anchor=west,inner sep=2pt]{\footnotesize$s_1$};
        \draw[->,>=stealth,thick](0.0,0.0)--(0.0,0.75)node[anchor=south,inner sep=2pt]{\footnotesize$s_2$};
        \draw[black,thick](-3,1.5)--(3,1.5);
        \draw[black,thick](-3,-1.5)--(3,-1.5);
        \draw[black,thick](1.5,-3)--(1.5,3);
        \draw[black,thick](-1.5,-3)--(-1.5,3);
        \draw[black,thick](1.2426,3)--(3,1.2426);
        \draw[black,thick](0.34835,-3)--(-3,0.34835);
        \draw[black,line width=2pt](-1.5,1.5)--(1.5,1.5);
        \draw[black,line width=2pt](-1.5,1.5)--(-1.5,-1.1517);
        \draw[black,line width=2pt](1.5,1.5)--(1.5,-1.5);
        \draw[black,line width=2pt](1.5,-1.5)--(-1.1517,-1.5);
        \draw[black,line width=2pt](-1.1517,-1.5)--(-1.5,-1.1517);
        \filldraw[black](1.5,1.5)circle(2pt)node[anchor=south west,inner sep=2pt]{\footnotesize$1$};
        \filldraw[black](-1.5,1.5)circle(2pt)node[anchor=south east,inner sep=2pt]{\footnotesize$2$};
        \filldraw[black](-1.5,-1.5)circle(2pt)node[anchor=north east,inner sep=2pt]{\footnotesize$3$};
        \filldraw[black](1.5,-1.5)circle(2pt)node[anchor=north west,inner sep=2pt]{\footnotesize$4$};
        \filldraw[black](1.5,2.7426)circle(2pt)node[anchor=north east,inner sep=2pt]{\footnotesize$5$};
        \filldraw[black](2.7426,1.5)circle(2pt)node[anchor=north east,inner sep=2pt]{\footnotesize$6$};
        \filldraw[black](-1.5,-1.1517)circle(2pt)node[anchor=south west,inner sep=2pt]{\footnotesize$7$};
        \filldraw[black](-1.1517,-1.5)circle(2pt)node[anchor=south west,inner sep=2pt]{\footnotesize$8$};
        \draw [->,>=stealth,dashed,thick](1.5,-1)coordinate[](a1)--(2.25,-1)coordinate[](a2) node[anchor=west,inner sep=2pt]{\footnotesize$\{p\}^a$};
        \draw [->,>=stealth,dashed,thick](-1.5,1)coordinate[](a1m)--(-2.25,1)coordinate[](a2m) node[anchor=east,inner sep=2pt]{\footnotesize$\{p\}^{a^-}$};
        \draw [->,>=stealth,dashed,thick](-1,1.5)coordinate[](b1)--(-1,2.25)coordinate[](b2) node[anchor=south,inner sep=2pt]{\footnotesize$\{p\}^b$};
        \draw [->,>=stealth,dashed,thick](1,-1.5)coordinate[](b1m)--(1,-2.25)coordinate[](b2m) node[anchor=north,inner sep=2pt]{\footnotesize$\{p\}^{b^-}$};
        \draw [->,>=stealth,dashed,thick](2.1213,2.1213)coordinate[](c1)--(2.6517,2.6517)coordinate[](c2)node[anchor=south west,inner sep=2pt]{\footnotesize$\{p\}^c$};
        \draw [->,>=stealth,dashed,thick](-2.1213+1.58,-2.1213)coordinate[](c1m)--(-2.6517+1.58,-2.6517)coordinate[](c2m)node[anchor=north,inner sep=2pt]{\footnotesize$\{p\}^{c^-}$}; 
    \end{tikzpicture}
    }
    \caption{Simplified two dimensional example of planes associated with slip conditions (solid lines) along with their normal vectors (dashed lines), and candidate vertices (dots, numbered) for the case of three possible slip systems---$a$, $b$, and $c$---plotted in deviatoric stress space. In this example, the critical resolved shear stress of slip system $c$ is chosen to be anisotropic between the positive and negative slip directions to demonstrate the effects on the single crystal yield surface calculation. The single crystal yield surface is highlighted (thick solid lines). Candidate vertices 9, 10, 11, and 12 are not shown for sake of simplicity}
    \label{fig:2dys_5}
\end{figure}
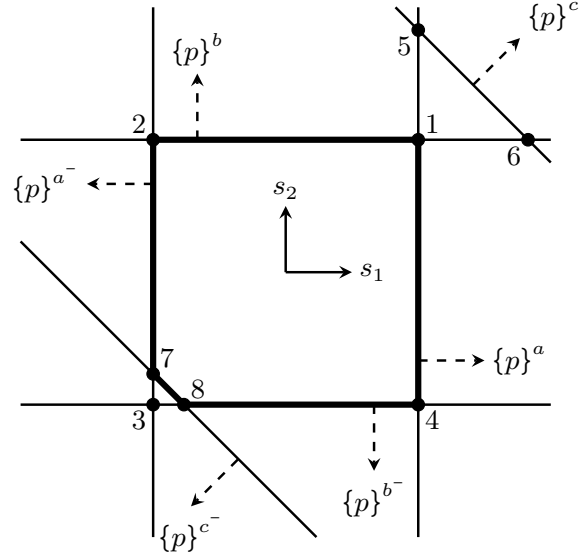

Finally, we consider a final case of strength anisotropy between the positive and negative slip directions for slip system $c$. In this case, we stipulate that $\left|\tau^{c^+}\right| = 2$ and $\left|\tau^{c^-}\right| = 1.25$, while slip systems $a$ and $b$ are held fixed, as before. We present a depiction of this in Figure~\ref{fig:2dys_5}, and note that vertices 1, 2, 4, 7, and 8 contribute to the single crystal yield surface. Interestingly, the symmetry of the single crystal yield surface is broken owing to the uneven strengths on slip system $c$.

\begin{figure}[H]
    \centering
    \subfigure[]{%
        \label{subfig:2d_h_matrix_a}
        \includegraphics[width=2in]{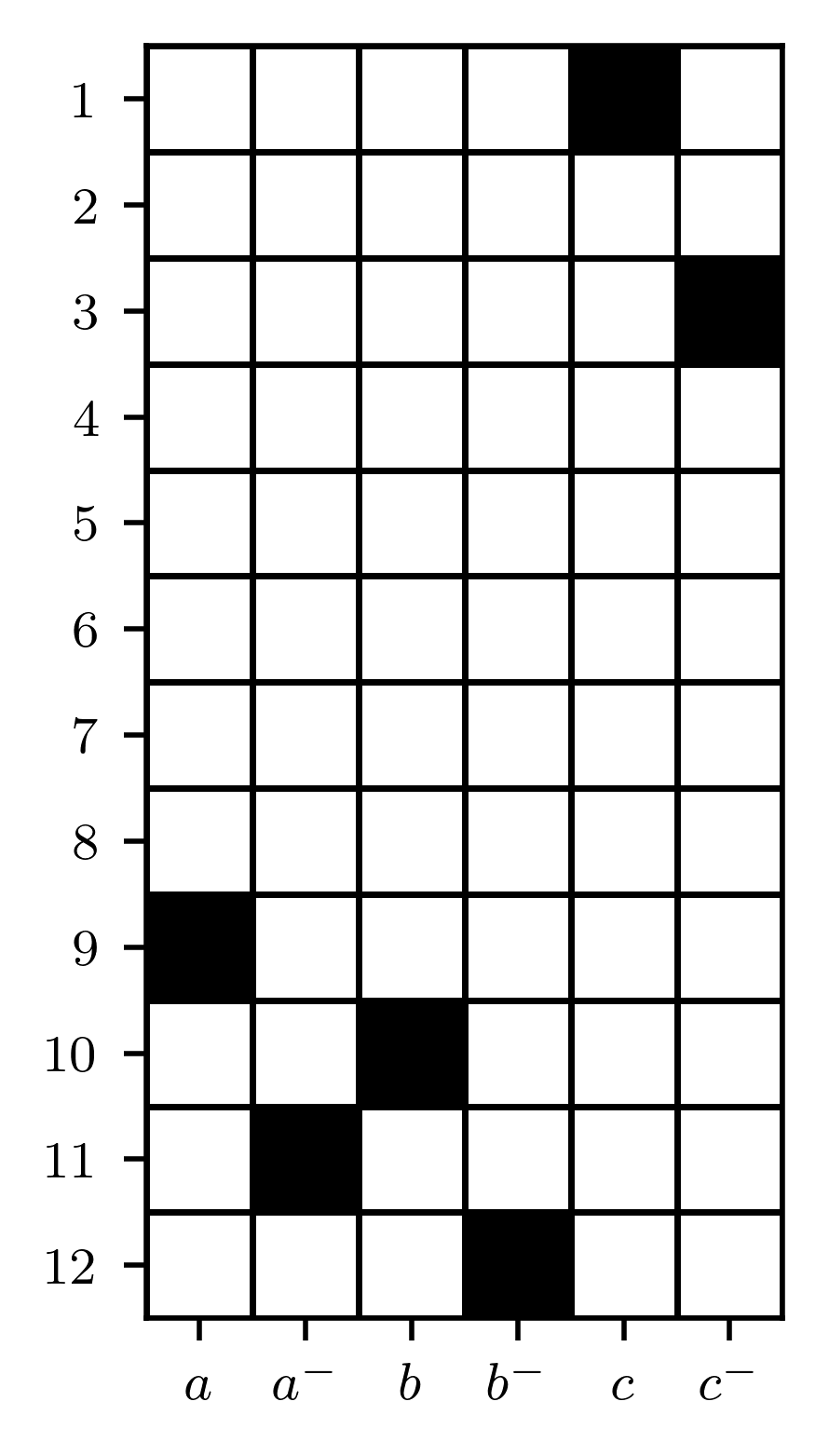}
    }
    \subfigure[]{%
        \label{subfig:2d_h_matrix_b}
        \includegraphics[width=2in]{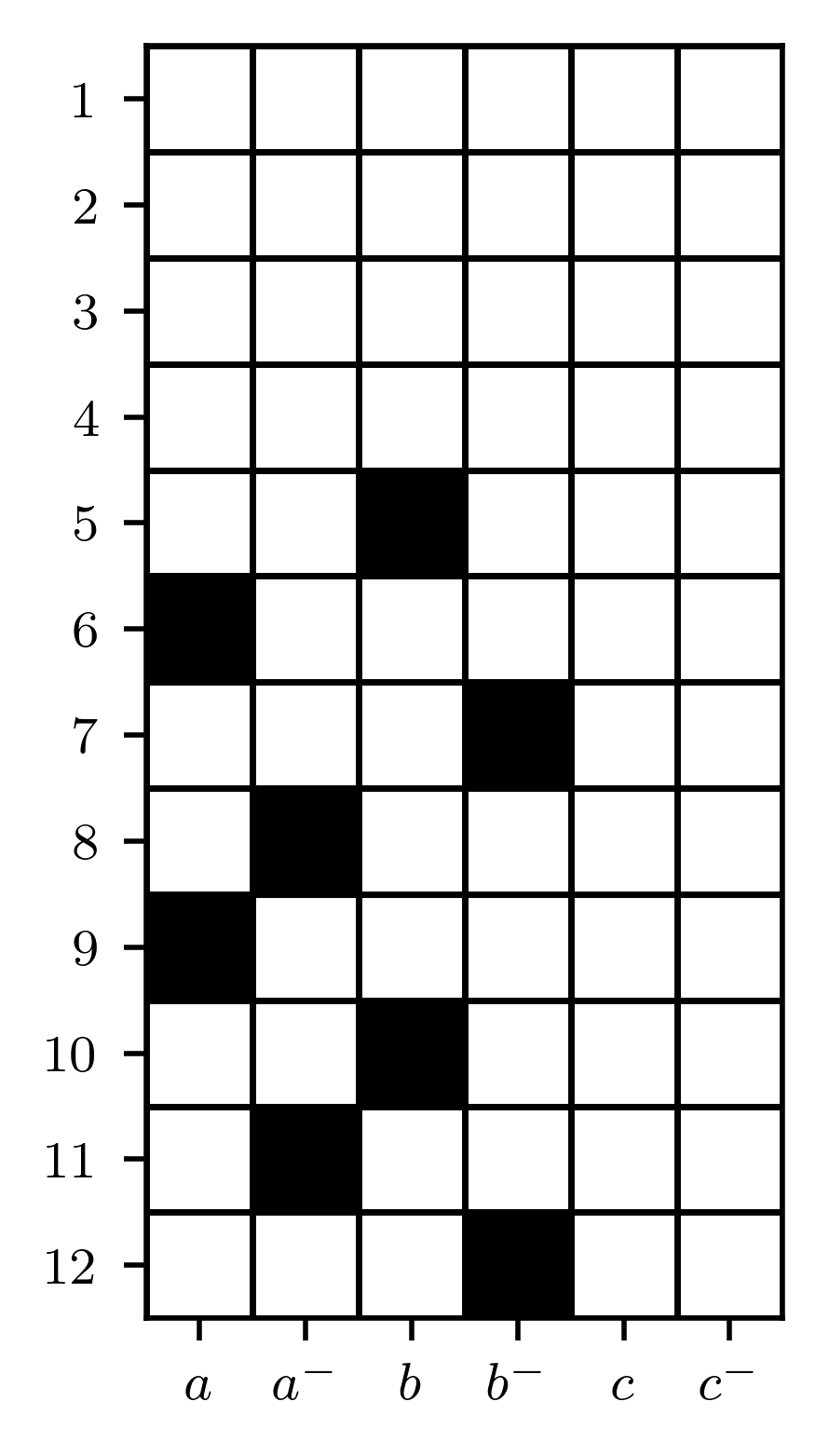}
    }
    \subfigure[]{%
        \label{subfig:2d_h_matrix_c}
        \includegraphics[width=2in]{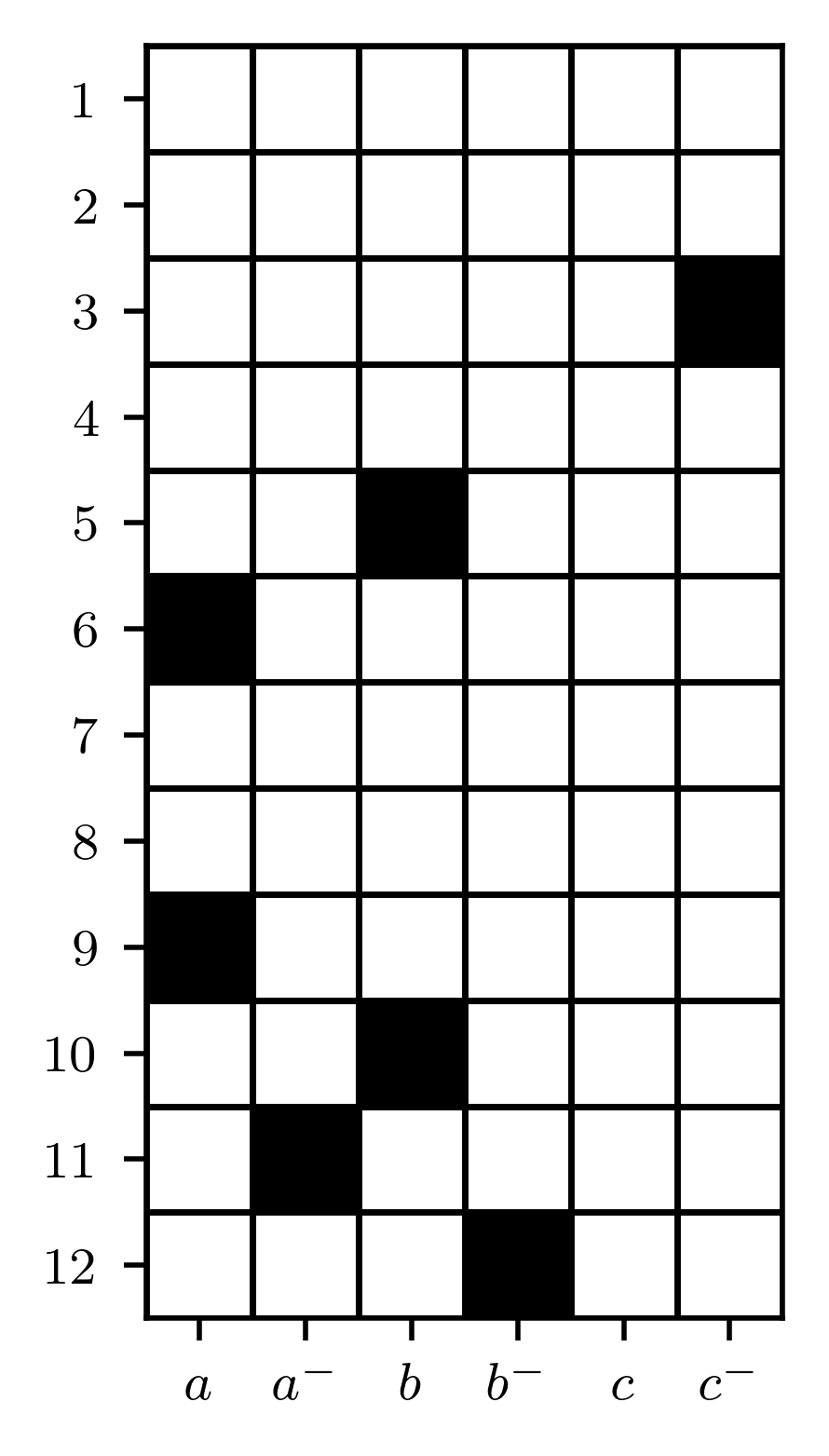}
    }
    \caption{Visualization of $h_{\alpha\beta}$ where black squares indicate an $h$ value greater than zero (i.e., candidate vertices which lie outside of a hyperplane) and white squares less than or equal to zero (i.e., candidate vertices which lie inside or on a hyperplane, respectively). Row indices and column indices correspond to the candidate vertex and hyperplane indexing, respectively, in Figure~\ref{fig:2dys_2}. Matrices are presented for ~\subref{subfig:2d_h_matrix_a} the simplified two dimensional example problem presented in Figure~\ref{fig:2dys_2},~\subref{subfig:2d_h_matrix_b} the problem presented in Figure~\ref{fig:2dys_4}, and~\subref{subfig:2d_h_matrix_c} the problem presented in Figure~\ref{fig:2dys_5}. Entirely white rows correspond to candidate vertices which are found to represent the vertices of the single crystal yield surface (i.e., the candidate vertex lies on our inside of every hyperplane).}
    \label{fig:2d_h_matrix}
\end{figure}

We further visualize the $h_{\alpha\beta}$ matrices for the problems demonstrated in Figures~\ref{fig:2dys_2},~\ref{fig:2dys_4}, and~\ref{fig:2dys_5} in Figures~\ref{subfig:2d_h_matrix_a},~\ref{subfig:2d_h_matrix_b} and~\ref{subfig:2d_h_matrix_c}, respectively. In these figure pairings, compare the rows of the $h_{\alpha\beta}$ with each entry less than or equal to zero against the vertices of the single crystal yield surface. Interestingly (beyond the primary function of identifying vertices by finding rows comprised entirely of zeroes), we note that the last two columns of Figure~\ref{subfig:2d_h_matrix_b} are comprised entirely of values less than or equal to zero, as all candidate vertices are on or inside the hyperplanes associated with slip systems $c^+$ and $c^-$, owing to their relatively high strength---i.e., no candidate vertices ``run afoul'' of these particular slip systems.

\section{Verification of the methodology via comparison to archival publications}

\subsection{FCC and BCC crystals (Bishop and Hill)}
\label{subsec:fcc_demonstration}

We now turn to a more realistic and relevant example of cubic crystals in $\mathbb{D}^5$. Bishop and Hill discussed the calculation of the vertices of the single crystal yield surface for FCC materials in depth in~\cite{Bishop1951}. Here, we attempt to reproduce these vertices in an effort to verify our method against the accepted results. We consider \hkl{111}\hkl<110> slip in FCC crystals (note that this is operationally identical to considering \hkl{110}\hkl<111> slip in BCC crystals) and assume a unit critical resolved shear stress, equal among the 12 slip systems in both positive and negative slip. We present the calculated vertices in Table~\ref{tab:fcc_verts} in the form of deviatoric stress vectors. The values presented here are equivalent to those calculated by Bishop and Hill (see: Section 4 of~\cite{Bishop1951}), considering the differences in normalizing factors and algebraic manipulations (described in Equation 10 of~\cite{Bishop1951}, compare to Equation~\ref{eq:svec} in this study).
\begin{table}[H]
    \centering
    \begin{tabular}{c c c c c | c}
        $\sqrt{\frac{1}{2}}\left(\sigma^\prime_{11} - \sigma^\prime_{22}\right)$ & $\sqrt{\frac{3}{2}}\sigma^\prime_{33}$ & $\sqrt{2}\sigma^\prime_{23}$ & $\sqrt{2}\sigma^\prime_{13}$ & $\sqrt{2}\sigma^\prime_{12}$ & Type~\cite{Bishop1951} \\
        \hline
        $\sqrt{3}$ & 1 & 0 & 0 & 0 & 1 \\
        0 & 0 & $\sqrt{12}$ & 0 & 0 & 2 \\ 
        $\sfrac{\sqrt{3}}{2}$ & $\sfrac{3}{2}$ & -$\sqrt{3}$ & 0 & 0 & 3 \\
        $\sfrac{\sqrt{3}}{2}$ & $\sfrac{1}{2}$ & $\sqrt{3}$ & 0 & $\sqrt{3}$ & 4 \\
        0 & 0 & $\sqrt{3}$ & $\sqrt{3}$ & $\sqrt{3}$ & 5 \\
    \end{tabular}
    \caption{5 of the 56 deviatoric stress vectors that describe the vertices of the single crystal yield surface for the case of an FCC material plastically deforming by \hkl{111}\hkl<110> slip or a BCC material plastically deforming by \hkl{110}\hkl<111> slip. 23 of the remaining stress vectors are symmetry transformations of these five (see:~\ref{sec:transform} for a description of vector transformation), and the other remaining 28 stress vectors are negatives of the other 28}
    \label{tab:fcc_verts}
\end{table}

\subsection{HCP crystals (\texorpdfstring{Tom{\'e}}{Tome} and Kocks)}
\label{subsec:hcp_demonstration}

We further demonstrate our method by considering the case covered by Tom{\'e} and Kocks~\cite{Tome1985}. To reproduce their results, we predict the vertices of the single crystal yield surface of zirconium crystals, with an idealized lattice ratio of $C=1.59$. For sake of making the comparison as simple as possible, we reproduce the (purely ``academic'') case of slip only on the pyramidal $c+a$ (\hkl{10-11}\hkl<11-2-3>) slip family with a unit critical resolved shear stress (see: Table 4 of~\cite{Tome1985} for their results, considering their choice of deviatoric stress vector construction in Equation 17 of~\cite{Tome1985}). We present the results of our calculations in Table~\ref{tab:hcp_verts} and note that the values we find are equivalent to those presented by Tom{\'e} and Kocks\footnote{We match the values presented by Tom{\'e} and Kocks exactly, with the exception of the $\sigma^\prime_{23}$ component of the third vertex. Tom{\'e} and Kocks report a coefficient of $-\sqrt{8}$ in Table 4 of~\cite{Tome1985} (which would require a coefficient of $-\sqrt{16}$ rather than $-\sqrt{24}$ in our presentation of stress vertices in Table~\ref{tab:hcp_verts} of this manuscript, since we scale our shear components by $\sqrt{2}$ relative to Tom{\'e} and Kocks). However, the stress state utilizing the reported coefficient does not produce slip on any pyramidal slip systems, and thus it cannot be a vertex. We contend that the value of $-\sqrt{8}$ presented by Tom{\'e} and Kocks is an error, and should actually be $-\sqrt{12}$, which indeed satisfies Schmid's law on six slip systems and (when scaled) matches the results found in this study.}. We use the same notation as in~\cite{Tome1985} for easing the translation of our representation for comparison, namely:
\begin{equation}
    \upbeta = \sqrt{\frac{4\left(1+C^2\right)\left(3+4C^2\right)}{3C^2}}.
\end{equation}
\begin{table}[H]
    \scriptsize
    \centering
    \begin{tabular}{c c c c c | c}
        \normalsize $\sqrt{\frac{1}{2}}\left(\sigma^\prime_{11} - \sigma^\prime_{22}\right)$ & \normalsize$\sqrt{\frac{3}{2}}\sigma^\prime_{33}$ & \normalsize$\sqrt{2}\sigma^\prime_{23}$ & \normalsize$\sqrt{2}\sigma^\prime_{13}$ & \normalsize$\sqrt{2}\sigma^\prime_{12}$ & \normalsize Type~\cite{Tome1985} \strut \\
        \hline
        0 & $-\sqrt{\frac{1}{6}}\upbeta$ & 0 & 0 & 0 & 1 \\
        $\sqrt{\frac{1}{2}}\upbeta$ & 0 & 0 & 0 & 0 & 2 \\
        $\sqrt{\frac{1}{2}}\left(\frac{2C^2-3}{8C^2-6}\right)\upbeta$ & $-\sqrt{\frac{1}{6}}\left(\frac{2C^2-3}{8C^2-6}\right)\upbeta$ & $-\sqrt{24}\left(\frac{C}{8C^2-6}\right)\upbeta$ & 0 & 0 & 3 \\
        $\sqrt{\frac{1}{2}}\left(\frac{1}{2C^2-1}\right)\upbeta$ & 0 & 0 & $-\sqrt{2}\left(\frac{C}{2C^2-1}\right)\upbeta$ & 0 & 4 \\
        $\sqrt{\frac{1}{2}}\left(\frac{4C^2-4}{6C^2-5}\right)\upbeta$ & $\sqrt{\frac{1}{6}}\left(\frac{2C^2-3}{6C^2-5}\right)\upbeta$ & 0 & $-\sqrt{2}\left(\frac{2C}{6C^2-5}\right)\upbeta$ & 0 & 5 \\
        $\sqrt{\frac{1}{2}}\left(\frac{4C^4-8C^2+3}{12C^4-8C^2-3}\right)\upbeta$ & -$\sqrt{\frac{1}{6}}\left(\frac{4C^4-8C^2+3}{12C^4-8C^2-3}\right)\upbeta$ & $-\sqrt{6}\left(\frac{2C^3-C}{12C^4-8C^2-3}\right)\upbeta$ & $\sqrt{2}\left(\frac{2C^3-3C}{12C^4-8C^2-3}\right)\upbeta$ & $\frac{\sqrt{6}}{2}\left(\frac{4C^4-3}{12C^4-8C^2-3}\right)\upbeta$ & 6 \\
        $\sqrt{\frac{1}{2}}\left(\frac{4C^4-6C^2}{12C^4-3}\right)\upbeta$ & $-\sqrt{\frac{1}{6}}\left(\frac{4C^4-8C^2+3}{12C^4-3}\right)\upbeta$ & $-\sqrt{6}\left(\frac{2C^3+C}{12C^4-3}\right)\upbeta$ & $\sqrt{2}\left(\frac{2C^3-3C}{12C^4-3}\right)\upbeta$ & $\sqrt{6}\left(\frac{2C^4+C^2}{12C^4-3}\right)\upbeta$ & 7 \\
    \end{tabular}
    \caption{Seven of the 92 deviatoric stress vectors that describe the vertices of the single crystal yield surface for the case of an HCP material with a lattice ratio of $C=1.59$ and plastically deforming solely by \hkl{10-11}\hkl<11-2-3> slip. 39 of the remaining stress vectors are symmetry transformations of these seven (see:~\ref{sec:transform} for a description of vector transformation), and the other remaining 46 stress vectors are the negatives of the other 46.}
    \label{tab:hcp_verts}
\end{table}

\section{Extension to more complex cubic and hexagonal systems}
\label{sec:novel_predictions}

\subsection{BCC crystals with multiple slip planes sharing the same slip direction}
\label{subsubsec:bcc_with_multiple}

While commonly restricted to the \hkl{1 1 0}\hkl<1 1 1> slip family, slip in BCC crystals has been observed to occur on various slip families over a wide range of temperatures, and so-called ``pencil glide''~\cite{Piehler1971} in BCC crystals---or slip associated with any family with a \hkl<1 1 1> slip direction---has been proposed to be a more accurate way to describe plastic deformation in BCC crystals. Consequently, we here consider the case of a BCC crystal which can plastically deform on the \hkl{1 1 2}\hkl<1 1 1> slip family in addition to the \hkl{1 1 0}\hkl<1 1 1> slip family (i.e., 24 slip systems in total) to judge the effects of the inclusion of further slip families associated with \hkl<1 1 1> slip on the topology of the single crystal yield surface. We concede that the consideration of these two slip systems does not fully describe the conditions theorized by pencil glide, but this case serves simply to motivate the inclusion of further slip families in a system that is commonly modeled via only one (an interesting study on the consideration of all possible planes is found in~\cite{Piehler1971}). We apply a unit strength to the \hkl{1 1 0} slip family, and vary the strength on the \hkl{1 1 2} slip family relative to the strength on the \hkl{1 1 0} slip family. Consequently, we will refer to the ratio of the familial slip strengths, or $\sfrac{\tau^{\hkl{1 1 2}}}{\tau^{\hkl{1 1 0}}}$. We do not modify the relative strengths to model any actual physical evolution of strengths, but rather to give a sense of the topology of the single crystal yield surface as a function of $\sfrac{\tau^{\hkl{1 1 2}}}{\tau^{\hkl{1 1 0}}}$.

\begin{figure}[H]
    \centering
    \includegraphics[width=3.5in]{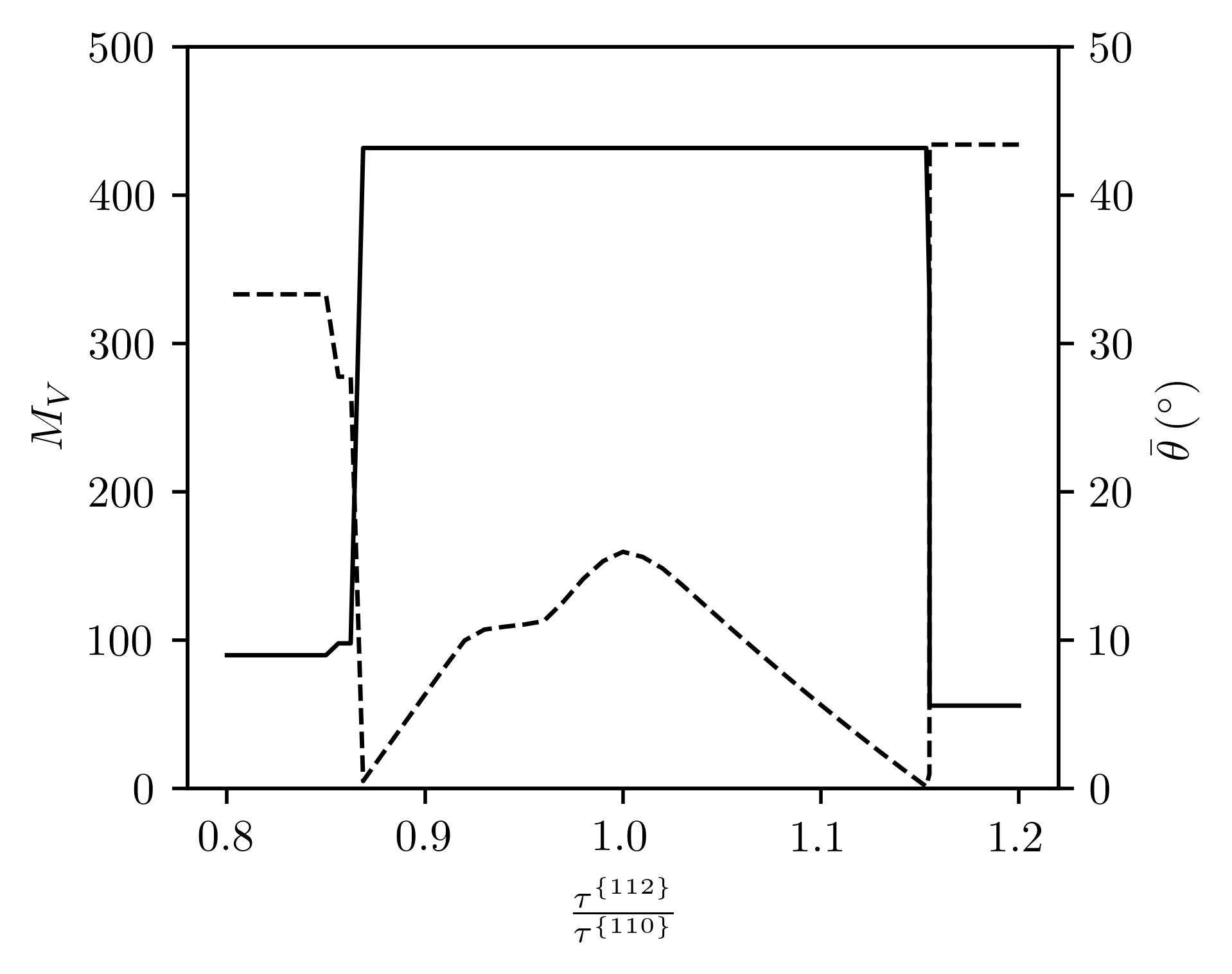}
    \caption{Evolution of the number of vertices, $M_V$, (solid line, left vertical axis) and the average angle to the nearest neighbor, $\bar{\theta}$, (dashed line, right vertical axis) as a function of the slip strength ratio, $\sfrac{\tau^{\hkl{1 1 2}}}{\tau^{\hkl{1 1 0}}}$. The left hand side of the plot (approximately $\sfrac{\tau^{\hkl{1 1 2}}}{\tau^{\hkl{1 1 0}}} < 0.85$) corresponds to a single crystal yield surface comprised purely of hyperplanes associated with \hkl{1 1 2}\hkl<1 1 1> slip systems, while the right hand side of the plot (approximately $\sfrac{\tau^{\hkl{1 1 2}}}{\tau^{\hkl{1 1 0}}} > 1.15$) corresponds to a single crystal yield surface comprised purely of hyperplanes associated with \hkl{1 1 0}\hkl<1 1 1> slip systems, while values in between are associated with single crystal yield surfaces with contribution from both slip families.}
    \label{fig:bcc_112_trends}
\end{figure}

Per the example in $\mathbb{D}^2$ presented in Figures~\ref{fig:2dys_2} and~\ref{fig:2dys_4}, we are already acquainted with the consequences of relative slip strengths between slip systems in terms of the number of vertices, $M_V$, and the average spacing between vertices and their nearest neighbor, or $\bar{\theta}$ (Equation~\ref{eq:bartheta}). We contend similar behavior occurs here in $\mathbb{D}^5$, where there exists a critical strength for the \hkl{1 1 2} slip family at which point it will no longer contribute to the single crystal yield surface (or, alternatively, a critical strength at which the \hkl{1 1 0} slip family no longer contributes). To this end, we vary $\sfrac{\tau^{\hkl{1 1 2}}}{\tau^{\hkl{1 1 0}}}$ to find these critical points. We find that the \hkl{1 1 2} slip family does not contribute to the single crystal yield surface when its strength is roughly 15\% greater than the \hkl{1 1 0} slip family (i.e., the vertices calculated are identical to those found in Section~\ref{subsec:fcc_demonstration}), and that the \hkl{1 1 2} slip family is the sole contributor to the single crystal yield surface when its strength is roughly 15\% less than the \hkl{1 1 0} slip family. We quantify the number of single crystal yield surface vertices, and the average angle to the nearest neighbor for all vertices, $\bar{\theta}$, both as a function of $\sfrac{\tau^{\hkl{1 1 2}}}{\tau^{\hkl{1 1 0}}}$ and present these data in Figure~\ref{fig:bcc_112_trends}. 

We note the rapid increase in number of vertices when the \hkl{1 1 2} slip family begins to contribute to the single crystal yield surface jointly with the \hkl{1 1 0} slip family (i.e., when $\sfrac{\tau^{\hkl{1 1 2}}}{\tau^{\hkl{1 1 0}}}$ falls below approximately $1.15$). This is accompanied by a sharp decrease in $\bar{\theta}$, per the phenomenon described previously for the two dimensional case in Section~\ref{subsubsec:three_slip_systems} regarding the introduction of new vertices and the angle to the nearest neighbor. As the hyperplanes associated with the \hkl{1 1 2} slip family begin to contribute to the single crystal yield surface, they first enter (or clip) the existing yield surface (comprised of vertices associated only with the \hkl{1 1 0} slip family) at the point of the existing vertices, thus creating multiple new vertices near the locations of the previous vertices. These new vertices are tightly clustered, and thus the angle to the nearest neighbor for these new vertices is low, significantly decreasing the average across all vertices (we note that this is primarily driven by the tight clustering of vertices, though we still expect these clusters to be roughly located at the approximate location of the clipped \hkl{1 1 0} vertices). As the strength of the \hkl{1 1 2} slip family is further decreased, the number of vertices stays constant but $\bar{\theta}$ begins to increase, and then decreases again as the hyperplanes associated with the \hkl{1 1 2} slip family begin to annihilate the remaining vertices associated with the \hkl{1 1 0} slip family. In other words, the portions of hyperplanes associated with the \hkl{1 1 0} slip family that contribute to the single crystal yield surface will become increasingly smaller as those associated with \hkl{1 1 2} slip dominate, and thus the collection of vertices that define these waning \hkl{1 1 0} facets will begin to converge before disappearing from the single crystal yield surface when it becomes entirely comprised of planes associated with \hkl{1 1 2} slip.

\subsection{BCC crystals with unequal slip system strengths within a single family}
\label{subsec:bcc_with_anisotropy}

We now consider the case in which slip systems in the same family of Schmid tensors have unequal strengths (i.e., rather than the strength disparity explored between differing slip families, as discussed in Section~\ref{subsubsec:bcc_with_multiple}). Having one or multiple slip systems stronger or weaker than the others in a given slip family could conceivably have several origins/causes arising from, for example, anisotropic or latent hardening~\cite{Kocks1964b}, or due to environment effects such as the formation of dislocation channels due to radiation exposure. Overall, these phenomena are not of principal interest for this study, but simply serve to motivate our methodology which provides the means to deal with material systems that lack complete symmetries as had been done in the analyses of Bishop and Hill for FCC (Section~\ref{subsec:fcc_demonstration}) and Tom{\'e} and Kocks for HCP (Section~\ref{subsec:hcp_demonstration}). Indeed, the example in Section~\ref{subsubsec:bcc_with_multiple} is expected to retain a high degree of symmetry, as every slip system in a given family is given the same strength (in both the positive and negative slip direction).

Here, we aim to demonstrate the consequence of a breakdown in symmetry. For this, we consider BCC crystals deforming solely by \hkl{1 1 0}\hkl<1 1 1> slip, and alter the strengths such that any one of the slip systems is weaker than the rest (or, alternatively: where 11 slip systems are stronger than one). The choice of which slip system to hold fixed (and thus those to strengthen) is arbitrary owing to the symmetry otherwise present (i.e., the same single crystal yield surface will be predicted no matter what slip system is held fixed versus which are strengthened, albeit in a rotated configuration in deviatoric stress space compared to if other slip systems are chosen as the weak slip system).

Recall, when all slip systems are equal strength (see: Section~\ref{subsec:fcc_demonstration}), then the single crystal yield surface is comprised of $M_V=56$ vertices (Table~\ref{tab:fcc_verts}), with an average angle to the nearest neighbor of $\bar{\theta}=\SI{43.43}{\degree}$. When we alter the slip system strengths such that 11 of the slip systems are 5\% stronger than one of the slip systems, then the predicted single crystal yield surface is comprised of $M_V=116$ vertices, with an average angle to the nearest neighbor of $\bar{\theta}=\SI{11.87}{\degree}$. In other words, moving all of the hyperplanes (except one) farther from the origin causes an increase in the number of hyperplane intersections that contribute to the single crystal yield surface, though the introduction of those new vertices overall decreases the average spacing, indicating that they are highly clustered on or around the facet of the yield surface associated with the weak slip system. Interestingly, however, $M_V$ does not change as a function of the strength disparity between the weak and the strong slip systems, though the value of $\bar{\theta}$ does decrease. This indicates that despite moving 11 of the 12 hyperplanes farther from the origin, the one hyperplane is still necessary to satisfy slip conditions in a certain stress direction (i.e., to ``close'' the single crystal yield surface), and the decrease in $\bar{\theta}$ is consistent with angles to the nearest neighbor for the vertices associated with the weak slip system becoming more acute.

\subsection{HCP crystals with multiple types of pyramidal slip}
\label{subsubsec:hcp_with_a}

\begin{figure}[H]
    \centering
    \includegraphics[width=3.5in]{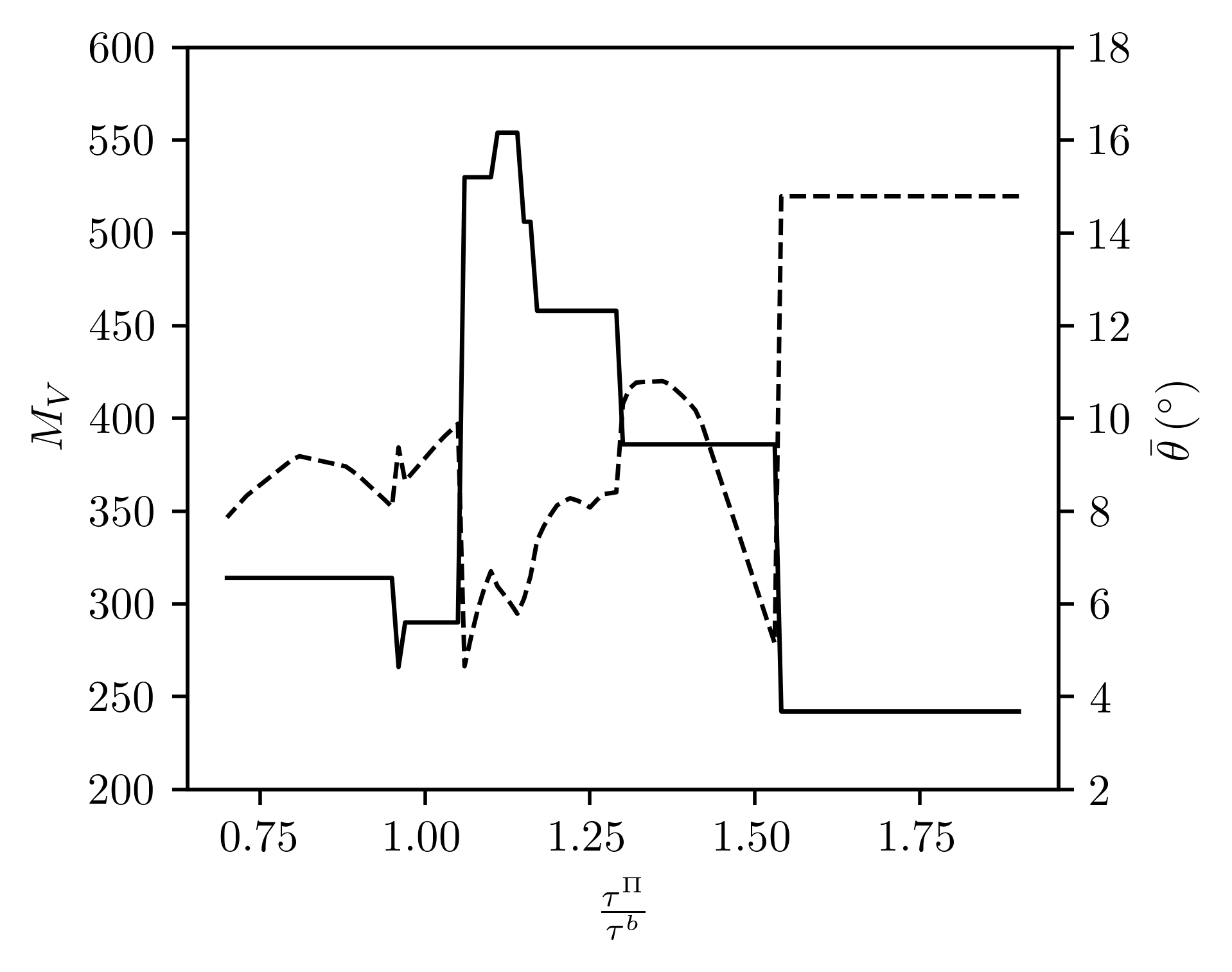}
    \caption{Evolution of the number of vertices, $M_V$, (solid line, left vertical axis) and the average angle to the nearest neighbor, $\bar{\theta}$, (dashed line, right vertical axis) as a function of the slip strength ratio, $\sfrac{\tau^{\Pi}}{\tau^{b}}$.}
    \label{fig:hcp_pyr_a_trends}
\end{figure}

Often, slip in HCP crystals is assumed to be restricted to the basal (\hkl{0 0 0 1}\hkl<1 1 -2 0>, denoted by the index $b$), prismatic (\hkl{1 0 -1 0}\hkl<1 1 -2 0>, denoted by the index $p$), pyramidal $c+a$ (\hkl{1 0 -1 1}\hkl<1 1 -2 -3>, denoted by the index $\pi$) slip families, though there is debate as to whether other slip families---such as the pyramidal $a$ (\hkl{1 0 -1 1}\hkl<1 1 -2 0>, denoted by the index $\Pi$) slip family---contribute appreciably to the plastic deformation response of HCP crystals, especially early in the development of plasticity. This presents an interesting problem to be addressed by the topology of single crystal yield surface, as the presence (or lack) of hyperplanes associated with the $\Pi$ family on the single crystal yield surface as a function of its relative strength highlights its relative availability to accommodate plastic deformation. Consequently, we consider the case of an HCP crystal which can plastically deform on the traditional $b$, $p$, and $\pi$ slip families, and additionally the $\Pi$ slip family. We choose $\upalpha$ titanium for this case, owing to the high degree of certainty in the determination of the slip system strengths for the basal, prismatic, and pyramidal $c+a$ slip systems: where for a unit strength of the basal slip system, we assume $\sfrac{\tau^p}{\tau^b}=1.2$ and $\sfrac{\tau^{\pi}}{\tau^{b}}=1.7$, and hold these relative strengths fixed. We assume a lattice ratio of $C=1.587$. The question turns to what effect the pyramidal $a$ slip system has on the single crystal yield surface, and as such we vary the ratio of $\sfrac{\tau^{\Pi}}{\tau^b}$ in a similar way to the analysis performed in Section~\ref{subsubsec:bcc_with_multiple}: again, we do not modify the relative strength of the pyramidal $a$ slip family to model any actual evolution of strengths, but rather to give a sense of the topology of the single crystal yield surface as a function of the choice of relative strengths. We present these results in Figure~\ref{fig:hcp_pyr_a_trends}.

We first caution that the behavior shown in Figure~\ref{fig:hcp_pyr_a_trends} is a function of both the choice of the lattice ratio, $C$, as well as the slip strengths of the $b$, $p$, and $\pi$ slip families, and as such should not be considered absolutely general. With that in mind, we note that there exists a large variation in the topology of the single crystal yield surface as a function of $\sfrac{\tau^{\Pi}}{\tau^{b}}$, both in terms of the number of vertices that comprise the single crystal yield surface ($M_V$), as well as the average spacing between vertices and their nearest neighbor, $\bar{\theta}$. When $\sfrac{\tau^{\Pi}}{\tau^{b}}$ is greater than approximately $1.55$, the $\Pi$ slip family does not contribute to the single crystal yield surface, and the prediction becomes the same as if only the $b$, $p$, and $\pi$ slip families are considered. As the hyperplanes associated with the $\Pi$ slip family begin to contribute to the single crystal yield surface (i.e., when the relative strength drops below 1.55), the value of $\bar{\theta}$ drops and the value of $M_V$ rises, both considerably, as the existing single crystal yield surface vertices are clipped and multiply into clusters of new vertices. Trends and interpretation are similar to those in Section~\ref{subsubsec:bcc_with_multiple} until the hyperplanes associated $\Pi$ system begins to interact more considerably with the hyperplanes associated with the lower strength $b$ and $p$ slip systems (approximately $\sfrac{\tau^{\Pi}}{\tau^{b}}<1.3$), at which point $M_V$ and $\bar{\theta}$ fluctuate considerably. This persists until $\sfrac{\tau^{\Pi}}{\tau^{b}}$ falls below approximately 0.9, at which point $M_V$ stabilizes and $\bar{\theta}$ decreases at a steady rate (a trend expected to continue as the hyperplanes associated with the $\Pi$ slip family move increasingly closer to the origin). Overall, we attribute the large degree of variability of the single crystal yield surface as a function of $\sfrac{\tau^{\Pi}}{\tau^{b}}$ to the presence of multiple slip families with unequal strengths. The yield surface comprised of $b$, $p$, and $\pi$ slip families is already highly faceted (comprised of hundreds of vertices), and as the hyperplanes associated with $\Pi$ move closer to the origin, they clip the hyperplanes and vertices associated with $b$, $p$, and $\pi$ slip families at different points, leading to relatively complex behavior, in comparison to that displayed in Figure~\ref{fig:bcc_112_trends} for \hkl{1 1 2} slip in BCC crystals.

\section{Conclusion}
\label{sec:conclusion}

In this study, we have presented a generalized calculation of the rate independent single crystal yield surface. This method considers arbitrary crystal symmetry, and allows for single crystal yield surface asymmetry via the consideration of inter- and intra-slip family strength anisotropy as well as strength anisotropy between the positive and negative slip directions, thus considering more complex asymmetric single crystal yield surface topologies not covered by archival publications. We presented two dimensional demonstrations to visualize the calculation of the single crystal yield surface vertices and the consequences of choice of slip systems, relative strengths, and the collapse of symmetry in various forms, as well as re-calculations of the vertices presented in archival publications for verification purposes (and indeed, highlight presumed errors in one of these publications). Finally, we presented novel single crystal yield surface calculations, and inspected metrics quantifying the evolution of the topology of the single crystal yield surface---namely the number of vertices and the average spacing to the nearest neighbor---demonstrating the rapid predictive abilities of the open-source code supplied with this study.

\section*{Acknowledgments}
\label{sec:acknowledgments}

MK would like to thank his doctoral students, Ezra Mengiste and Lloyd van Wees, for assistance with plotting as well as proofing and editing the manuscript, and further the University of Alabama for providing computational resources used in this study. We would further like to thank Prof. Darren Pagan for helpful comments during drafting of this article.

\section*{Data Availability}
\label{sec:data}

A computational implementation of this method in the Python language is hosted on GitHub at \url{https://github.com/mkasemer/scys} at time of publication, with an intent to maintain for at least five years after publication. This piece of code is disseminated free and open source, pursuant to the GNU General Public License, Version 3.

\begin{appendices}
\setcounter{figure}{0}
\renewcommand\thefigure{A.\arabic{figure}}
\addcontentsline{toc}{section}{Appendix}

\section{Deviatoric vector transformation}
\label{sec:transform}

In this paper, we perform all calculations and represent stress in the crystal frame. It may be advantageous or necessary to transform the stress for various reasons, such as the necessity to perform calculations in a secondary frame, or to apply crystal symmetry operators for the calculation of single crystal yield surface vertices associated with the various symmetric slip systems, as described in Section~\ref{subsec:hcp_demonstration}. We here demonstrate the transformation of a deviatoric stress vector from the crystal frame to a secondary frame.

We can perform a typical  tensor transformation, shown here for the deviatoric component of the Cauchy stress tensor:
\begin{equation}
    \label{eq:tenstrans}
    \left(\sigma_{mn}^\prime\right)^S = a_{mi} a_{nj} \left(\sigma_{ij}^\prime\right)^C ,
\end{equation}
where $C$ refers to the crystal frame and $S$ to a secondary frame, and $a$, the transformation matrix, is defined as:
\begin{equation}
    \label{eq:Rtrans}
    a_{ij} = e_j^C e_i^S,
\end{equation}
where $a_{ij}$ is constructed from the basis vectors, $e$, in each frame, and here transforms from the crystal frame to the secondary frame.

However, in this paper we are often considering symmetric, deviatoric tensors in vector form, such as the deviatoric stress vector in Equation~\ref{eq:svec}. To transform these vectors to a secondary frame, we wish to find a more compact transformation via a matrix $\left[Q\right]$, or:
\begin{equation}
    \{s\}^S = \left[Q\right] \{s\}^C,
\end{equation}

To begin, we convert the tensor transformation in Equation~\ref{eq:tenstrans} to a matrix operation utilizing Voigt notation:
\begin{equation}
    \{\sigma^\prime\}^S = \left[K\right]\{\sigma^\prime\}^C ,
\end{equation}
which takes the following form, utilizing the values obtained from Equation~\ref{eq:tenstrans}:
\footnotesize
\begin{equation}
    \begin{Bmatrix}
        \sigma_{11}^\prime \\
        \sigma_{22}^\prime \\
        \sigma_{33}^\prime \\
        \sigma_{23}^\prime \\
        \sigma_{13}^\prime \\
        \sigma_{12}^\prime
    \end{Bmatrix}^S
    = 
    \begin{bmatrix}
        a_{11}^2 & a_{12}^2 & a_{13}^2 & 2a_{12}a_{13} & 2a_{11}a_{13} & 2a_{11}a_{12} \\
        a_{21}^2 & a_{22}^2 & a_{23}^2 & 2a_{22}a_{23} & 2a_{21}a_{23}& 2a_{21}a_{22}  \\
        a_{31}^2 & a_{32}^2 & a_{33}^2 & 2a_{32}a_{33} & 2a_{31}a_{33}& 2a_{31}a_{32}  \\
        a_{21}a_{31} & a_{22}a_{32} & a_{23}a_{33} & a_{22}a_{33}+a_{23}a_{32} & a_{21}a_{33}+a_{31}a_{23} & a_{21}a_{32}+a_{22}a_{31} \\
        a_{11}a_{31} & a_{12}a_{32} & a_{13}a_{33} & a_{12}a_{33}+a_{13}a_{32} & a_{11}a_{33}+a_{13}a_{31} & a_{11}a_{32}+a_{12}a_{31} \\
        a_{11}a_{21} & a_{12}a_{22} & a_{13}a_{23} & a_{12}a_{23}+a_{13}a_{22} & a_{11}a_{23}+a_{13}a_{21} & a_{11}a_{22}+a_{12}a_{21} \\
    \end{bmatrix}
    \begin{Bmatrix}
        \sigma^\prime_{11} \\
        \sigma^\prime_{22} \\
        \sigma^\prime_{33} \\
        \sigma^\prime_{23} \\
        \sigma^\prime_{13} \\
        \sigma^\prime_{12} 
    \end{Bmatrix}^C
\end{equation}
\normalsize

Next, we define a matrix that transforms a six dimensional stress vector in Voigt notation to a five dimensional deviatoric vector as defined in Equation~\ref{eq:svec}:
\begin{equation}
    \left[V\right] = 
    \begin{bmatrix}
        \frac{1}{\sqrt{2}} & -\frac{1}{\sqrt{2}} & 0 & 0 & 0 & 0 \\
        0 & 0 & \sqrt{\frac{3}{2}} & 0 & 0 & 0 \\
        0 & 0 & 0 & \sqrt{2} & 0 & 0 \\
        0 & 0 & 0 & 0 & \sqrt{2} & 0 \\
        0 & 0 & 0 & 0 & 0 & \sqrt{2}
    \end{bmatrix} ,
\end{equation}
as well as the inverse of that matrix, which transforms from a deviatoric five dimensional vector to the six dimensional Voigt form:
\begin{equation}
    \left[V\right]^{-1} = 
    \begin{bmatrix}
        \frac{1}{\sqrt{2}} & -\frac{1}{\sqrt{6}} & 0 & 0 & 0 \\
        -\frac{1}{\sqrt{2}} & -\frac{1}{\sqrt{6}} & 0 & 0 & 0 \\
        0 & \sqrt{\frac{2}{3}} & 0 & 0 & 0 \\
        0 & 0 & \frac{1}{\sqrt{2}} & 0 & 0 \\
        0 & 0 & 0 & \frac{1}{\sqrt{2}} & 0 \\
        0 & 0 & 0 & 0 & \frac{1}{\sqrt{2}} \\
    \end{bmatrix} .
\end{equation}

Finally, to find $\left[Q\right]$, we employ:
\begin{equation}
    \left[Q\right] = \left[V\right] \left[K\right] \left[V\right]^{-1}
\end{equation}
or:
\footnotesize
\begin{equation}
    \begin{bmatrix}
        \frac{1}{2}\left(a_{11}^2-a_{12}^2-a_{21}^2+a_{22}^2\right) & \frac{\sqrt{3}}{2}\left(a_{13}^2-a_{23}^2\right) & a_{12}a_{13}-a_{22}a_{23} & a_{11}a_{13}-a_{21}a_{23} & a_{11}a_{12}-a_{21}a_{22} \\
        
        \frac{\sqrt{3}}{2}\left(a_{31}^2-a_{32}^2\right) & \frac{3}{2}a_{33}^2-\frac{1}{2} & \sqrt{3}a_{32}a_{33} & \sqrt{3}a_{31}a_{33} & \sqrt{3}a_{31}a_{32} \\

        a_{21}a_{31}-a_{22}a_{32} & \sqrt{3}a_{23}a_{33} & a_{22}a_{33}+a_{23}a_{32} & a_{21}a_{33}+a_{31}a_{23} & a_{21}a_{32}+a_{22}a_{31} \\
        
        a_{11}a_{31}-a_{12}a_{32} & \sqrt{3}a_{13}a_{33} & a_{12}a_{33}+a_{13}a_{32} & a_{11}a_{33}+a_{13}a_{31} & a_{11}a_{32}+a_{12}a_{31} \\
        
        a_{11}a_{21}-a_{12}a_{22} & \sqrt{3}a_{13}a_{23} & a_{12}a_{23}+a_{13}a_{22} & a_{11}a_{23}+a_{13}a_{21} & a_{11}a_{22}+a_{12}a_{21} \\
        
    \end{bmatrix}
\end{equation}
\normalsize

\end{appendices}

\bibliography{bibliography}
\bibliographystyle{elsarticle-num}

\end{document}